\documentclass[twocolumn]{aastex631}

\newcommand\teff{$T_\mathrm{eff}$} 
\usepackage{enumitem}
\usepackage{natbib}
\usepackage{graphicx}
\usepackage[caption=false]{subfig}

\begin{document}

\title{Classifying Cool Dwarfs: Comprehensive Spectral Typing of Field and Peculiar Dwarfs Using Machine Learning}

\correspondingauthor{Tianxing Zhou}
\email{t4zhou@ucsd.edu}

\author[0000-0002-3211-2213]{Tianxing Zhou}
\affiliation{Department of Physics, University of California, San Diego, La Jolla, California 92093, USA}

\author[0000-0002-9807-5435]{Christopher A.~Theissen}
\affiliation{Department of Astronomy \& Astrophysics, University of California, San Diego, La Jolla, California 92093, USA}

\author[0000-0002-0424-8519]{S.~Jean~Feeser}
\affiliation{Department of Astronomy \& Astrophysics, 525 Davey Laboratory, The Pennsylvania State University, University Park, PA, 16802, USA}

\author[0000-0003-0562-1511]{William M.~J.~Best}
\affiliation{Department of Astronomy, University of Texas at Austin, 2515 Speedway C1400, Austin, TX 78712, USA}

\author[0000-0002-6523-9536]{Adam J.~Burgasser}
\affiliation{Department of Astronomy \& Astrophysics, University of California, San Diego, La Jolla, California 92093, USA}

\author[0000-0002-1821-0650]{Kelle L. Cruz}
\affiliation{Department of Physics and Astronomy, Hunter College, City University of New York, 695 Park Avenue, New York, NY 10065, USA}
\affiliation{Department of Physics, Graduate Center, City University of New York, 365 Fifth Avenue, New York, NY 10016, USA}
\affiliation{Department of Astrophysics, American Museum of Natural History, Central Park West at 79th Street, New York, NY 10024, USA}

\author[0009-0009-1521-0426]{Lexu Zhao}
\affiliation{Department of Physics, University of California, San Diego, La Jolla, California 92093, USA}
\affiliation{Department of Physics, University of Florida, Gainesville, Florida 32611, USA}



\begin{abstract}
Low-mass stars and brown dwarfs---spectral types (SpTs) M0 and later---play a significant role in studying stellar and substellar processes and demographics, reaching down to planetary-mass objects.
Currently, the classification of these sources remains heavily reliant on visual inspection of spectral features, equivalent width measurements, or narrow-/wide-band spectral indices. 
Recent advances in machine learning (ML) methods offer automated approaches for spectral typing, which \replaced{become}{are becoming} increasingly important as large spectroscopic surveys such as Gaia, SDSS, and SPHEREx generate datasets containing millions of spectra.
We investigate the application of ML in spectral type classification on low-resolution ($R \sim 120$) near-infrared spectra of M0--T9 dwarfs obtained with the SpeX instrument on \added{the} NASA Infrared Telescope Facility. We specifically aim to classify the gravity- and metallicity-dependent subclasses for late-type dwarfs. 
We used binned fluxes as input features and compared the efficacy of spectral type estimators built using Random Forest (RF), \deleted{K-Nearest Neighbor (KNN), and }Support Vector Machine (SVM) \added{, and K-Nearest Neighbor (KNN) }models.
We tested the influence of different normalizations
and analyzed the relative importance of different spectral regions for surface gravity and metallicity subclass classification.
Our best-performing model (using KNN) classifies 
95.5$\pm0.6$\% 
of sources \added{to} within $\pm$1 SpT, and assigns surface gravity and metallicity subclasses 
with 89.5$\pm$0.9\% 
accuracy. 
We test the dependence of signal-to-noise ratio on classification accuracy and find sources with SNR $\gtrsim$ 60 have $\gtrsim$ 95\% accuracy. We also find that \replaced{$J$-band}{$zy$-band} plays the most prominent role in the RF model, with FeH and TiO having the highest feature importance. 
\end{abstract}



\section{Introduction} 
\label{sec:intro}


Spectral classification is a cornerstone of astronomy. 
By organizing and analyzing spectra based on their 
atomic and molecular features, and overall morphology, we have expanded our understanding of stars, brown dwarfs, galaxies, and numerous other sources
(e.g., \citealt{1918AnHar..91....1C, 1943assw.book.....M, 1987ApJS...63..295V, 2005ARA&A..43..195K}). 
Spectral typing continues to be used as a way of categorizing and comparing observed sources, and even with the present-day 
use of sophisticated statistical and computational comparison methods, classification ``by-eye"\replaced{;}{,} i.e., direct comparison to standards or template spectra, is still the typical approach used in most stellar classification systems
\citep[e.g.,][]{2011AJ....141...97W, 2021ApJS..253....7K}.
As astronomical samples become larger and surveys attempt to 
maximize the utility of large catalog data, more streamlined spectral-typing methods are needed.
Low-mass stars and brown dwarfs, collectively ``cool dwarfs,'' are objects with masses $\lesssim0.5~M_\odot$ and effective temperatures {\teff} $\lesssim$ 4,000~K which comprise more than 70\% of the stellar population in the Milky Way \citep{2006AJ....132.2360H, 2018AJ....155..265H, 2010AJ....139.2679B, 2021A&A...650A.201R}.
These objects span the mass boundaries between H-fusing stars and non-H-fusing brown dwarfs ($M \sim$ 0.075~$M_\odot$; \citealt{2001RvMP...73..719B}), and between brown dwarf and non-fusing planetary-mass objects ($M \sim$ 0.013~$M_\odot$; \citealt{2000ARA&A..38..337C}), and hence sample multiple paths for formation and evolution.
These sources are excellent targets for terrestrial exoplanet discovery \citep[e.g.,][]{2015ApJ...807...45D, 2016ApJ...816...66B, 2017Natur.542..456G} and challenge atmosphere models due to their complicated molecular and weather-driven variability \citep[e.g.,][]{2003A&A...402..701B, 2020A&A...637A..38P, 2021ApJ...920...85M, 2021ApJ...923..269K, 2024ApJ...963...73M, 2024arXiv240200758M}. 

The spectral sequence for cool dwarfs 
follows spectral classes M, L, T, and Y, from highest to lowest effective temperature \citep{1991ApJS...77..417K, 1999ApJ...519..802K, burgasser:2006:1067, cushing:2011:50}. Over the past decades, the study of low-mass stars and brown dwarfs has grown thanks to the development of deep optical, near-infrared (NIR), and mid-infrared (MIR) surveys such as the Two Micron All Sky Survey \citep[2MASS;][]{2000AJ....120.1579Y}, the Sloan Digital Sky Survey, \citep[SDSS;][]{2006AJ....131.1163S}, the Wide-field Infrared Survey Explorer \citep[WISE;][]{2010AJ....140.1868W}, the UKIRT Infrared Deep Sky Survey \citep[UKIDSS;][]{2007MNRAS.379.1599L}, and the Panoramic Survey Telescope and Rapid Response System \citep[Pan-STARRS;][]{2016arXiv161205560C}. However, the discovery of red point sources in photometric surveys only provides a starting point for characterization, and follow-up spectroscopy is critical to fully characterize the physical properties of the reddest stellar/substellar objects.




Spectral typing of cool dwarfs involves a holistic treatment of each source, as \replaced{all}{many} physical properties of the source can influence its spectral features (e.g., temperature, surface gravity, metallicity, age).
Temperature-sensitive features form the basis of the MLTY spectral sequence, and include the deepening of molecular features through the M dwarf sequence, the conversion of metal oxides to metal hydrides and the formation of condensates in the L dwarf sequence, the emergence of CH$_4$ bands in the T dwarf sequence, and the emergence of strong NH$_3$ and shift of flux to \replaced{mid-infrared}{MIR} wavelengths in the Y dwarf sequence.
Gravity-sensitive features, such as VO and K~\textsc{i} absorption bands, 
provide a proxy for surface gravity, and thus the age of a cool dwarf \citep{1997A&A...327L..29M, 2004ApJ...600.1020M}. 
Studies have proposed classifying cool dwarfs into different gravity types, such as the four-category series $\delta$, $\gamma$, $\beta$, and $\alpha$ ranging from lowest to highest surface gravities \citep{2005ARA&A..43..195K, 2009AJ....137.3345C,2015ApJS..219...33G},
or the three-category series \textsc{vl-g}, \textsc{int-g}, and \textsc{fld-g} approximately aligned with ages $\lesssim$ 10~Myr, $\sim$10--100~Myr, and $\gtrsim$100~Myr, respectively
\replaced{\citet{2013ApJ...772...79A}}{\citep{2013ApJ...772...79A}}.
It should be noted that age and gravity do not always coincide, and that many objects with low surface gravity features are found to be high-velocity objects in the field population \citep[e.g.,][]{2016ApJS..225...10F, 2017ApJ...845..110B, 2019ApJ...886..131G}. 
Metallicity-sensitive features, such as the relative strengths of TiO and CaH bands used to distinguish M subdwarfs \citep{1997AJ....113..806G, 2007ApJ...669.1235L},
have been used to define metallicity classes of dwarfs (d), subdwarfs (sd), extreme subdwarfs (esd), and ultrasubdwarfs (usd), corresponding to $\zeta_{\text{TiO/CaH}} > 0.825$, $0.500 < \zeta_{\text{TiO/CaH}} \leq 0.825$, $0.200 < \zeta_{\text{TiO/CaH}} \leq 0.500$, and $\zeta_{\text{TiO/CaH}} \leq 0.200$, respectively.
This framework has been extended to L subdwarfs through their enhanced metal hydrides (e.g., FeH) and strong
collision-induced H$_2$ absorption (CIA H$_2$)
\citep{2007ApJ...657..494B,2019MNRAS.486.1260Z};
and most recently with T subdwarfs, which are being identified in increasing numbers in WISE data \citep[e.g.,][]{2020ApJ...898...77S,2020ApJ...899..123M,2024arXiv241101378B} and deep JWST fields \citep[e.g.,][]{2024ApJ...962..177B}.


Traditional spectral typing methods rely on the precise measurement of specific spectral features and are often limited by the signal-to-noise ratio (SNR) \citep{2003ApJ...593.1074G}. 
In addition, these methods usually involve visual inspection or direct quantitative comparison to spectral standards, processes that are both time-consuming and potentially prone to subjective biases, especially when dealing with the voluminous datasets generated by modern astronomical surveys. 
Machine Learning (ML) algorithms offer an alternative approach to classification \citep[e.g.,][]{2014MNRAS.437..968C, 2022ApJ...924..114A}.
As a subset of artificial intelligence (AI) techniques, ML encompasses algorithms and statistical methods that can identify complex, non-linear patterns in data, \added{improving}
\deleted{enhancing} efficiency \replaced{and reducing subjectivity}{in classification} \citep{Azevedo2024}. 
Studies of ML approaches for cool dwarfs have demonstrated an enhanced reliability in spectral typing, 
allowing for the robust identification of foreground brown dwarfs
in deep 
Hubble Space Telescope spectral data \citep[e.g.,][]{2022ApJ...924..114A}, inference of the physical properties of brown dwarfs based on empirical standards \citep[e.g.,][]{2022MNRAS.513..516F, 2022A&A...657A.129A},
identification of unresolved binaries from single spectra \citep{2023RNAAS...7...13D}, and
new forms of model-fitting and retrieval analyses \citep{2023ApJ...954...22L,2024A&A...690A.357L}. 



The goal of this study is to classify cool dwarfs in the spectral type (SpT) range M0--T9, along with peculiarity classes 
that encompass surface gravities (\textsc{vl-g}, \textsc{int-g}, \textsc{fld-g}) and metallicities (d/sd, sd, esd). 
To achieve this, we explore
three \added{supervised} ML algorithms: Random Forest (RF), Support Vector Machine (SVM), and K-Nearest Neighbors (KNN) \deleted{, evaluating their performance to determine the} \deleted{best classifier for cool dwarfs}. \added{In supervised learning, a model is trained on labeled data, enabling it to learn and recognize patterns from the input. It is important to note, however, that while this approach reduces subjectivity in the classification of new spectra by applying a consistent learned logic, it inherently carries the subjectivity of the ``ground truth” defined by the human-classified training set. We evaluate the performance of these models to determine the best classifier for cool dwarfs.}
%
The article is organized as follows. Section \ref{sec:Spectral Sample} describes the spectral data sample.
Section \ref{sec:Methods} decribes the supervised machine learning
algorithms we used to classify the spectra. 
Section \ref{sec:results} presents the results of our models and compares their performance. 
Section \ref{sec:Discussion} summarizes our study and discusses the possible implications of our approach.

\section{Spectral Sample} 
\label{sec:Spectral Sample}

\subsection{Spectral Standards} 
\label{subsec:Spectral Standards}

The spectral data analyzed in this study were obtained with the SpeX infrared spectrograph \citep{2003PASP..115..362R} mounted on the 3-m NASA Infrared Telescope Facility (IRTF) on Maunakea, Hawaii. 
We used low-resolution ($\lambda/\Delta\lambda \approx 120$) prism-mode data which covers the near-infrared (NIR) band of 0.8--2.5\,$\mu$m.
Spectra were procured from the 
SpeX Prism Library Analysis Toolkit (\texttt{SPLAT}\footnote{\url{https://github.com/aburgasser/splat}}; \citealt{2017ASInC..14....7B}), 
which encompasses sources spanning spectral types M0--T9 based on a combination of optical and NIR classifications, including representatives of low surface gravity and subsolar metallicity classes.

\added{From this collection, }We identified 70 spectral standards with this set, based on 
dwarf standards defined in \citet{2010ApJS..190..100K}, 
surface gravity standards defined in \citet{2013ApJ...772...79A}, \citet{2015ApJS..219...33G}, and \citet{2018AJ....155...34C} (C18 in short),
and subdwarf standards defined in \citet{burgasser:2004:l73} and \citet{greco:2019:182}, listed in Table~\ref{tbl:standards}. 
These standards have SNR ranging from 2.5--975, with an average SNR of 141 and a median of 69. \added{The representative SNR for each spectrum was computed using SPLAT \citep{2017ASInC..14....7B}, which computes the median signal-to-noise for all spectral pixels with flux values in the upper 50th percentile of the spectrum.}  

\added{These 70 high-quality spectra serve as the basis for generating a much larger synthetic training dataset of 70,000 spectra, a process detailed in Section~\ref{subsubsec:Building the Training set}.
Our set of standards is not comprehensive across all subtypes, 
as surface gravity standards are only \added{currently} defined for late-M and L dwarfs,
and subdwarf standards are only \added{currently} defined for M and early-L dwarfs. This approach allows us to build a robust training set that thoroughly samples the feature space around each spectral standard.}

In this study, the term ``spectral type" is used to specifically refer to NIR spectral type, even though M and L dwarfs are traditionally classified at optical wavelengths \citep{1991ApJS...77..417K, 1999ApJ...519..802K}. 
Spectral subtypes are rounded down to the nearest subtype to provide consistency across classes. This process introduces an inherent uncertainty of $\pm$1 subtypes, which we adopt as the accuracy limit of our analysis.

\startlongtable
\begin{deluxetable}{clccc}
\tablewidth{\textwidth}
\tabletypesize{\scriptsize} 
\tablecaption{Spectral Type Standards
\label{tbl:standards}}
\tablehead{
  \colhead{SpT} &
  \colhead{Name} &
  \colhead{$J_{\mathrm{2M}}$} &
  \colhead{$\mathrm{SNR}_{\mathrm{med}}$} &
  \colhead{Refs.}
}
\startdata
M0 & Gliese 270 & 7.18 & 685 & 1 \\
M1 & Gl424 & 6.31 & 658 & 1 \\
M2 & Gliese 91 & 6.96 & 193 & 1 \\
M3 & Gl752A & 5.58 & 895 & 1 \\
M4 & Gliese 213 & 7.12 & 975 & 1 \\
M5 & Wolf 47 & 8.61 & 273 & 1 \\
M6 & LHS 1375 & 9.87 & 133 & 1,\,3 \\
M7 & VB 8 & 9.78 & 344 & 2,\,3 \\
M8 & VB 10 & 9.91 & 366 & 2,\,4 \\
M9 & LHS 2924 & 11.99 & 228 & 2,\,5 \\
L0 & 2MASP J0345432+254023 & 14.00 & 59 & 6,\,7 \\
L1 & 2MASSW J2130446-084520 & 14.14 & 153 & 1,\,2 \\
L2 & Kelu-1 & 13.41 & 111 & 8,\,9 \\
L3 & 2MASSW J1506544+132106 & 13.37 & 70 & 10 \\
L4 & 2MASS J21580457-1550098 & 15.04 & 42 & 1,\,2 \\
L5 & SDSS J083506.16+195304.4 & 16.09 & 20 & 11 \\
L6 & 2MASSI J1010148-040649 & 15.51 & 58 & 12,\,13 \\
L7 & 2MASSI J0103320+193536 & 16.29 & 49 & 2,\,14 \\
L8 & 2MASSW J1632291+190441 & 15.87 & 30 & 8,\,15 \\
L9 & DENIS-P J0255-4700 & 13.25 & 51 & 15 \\
T0 & SDSS J120747.17+024424.8 & 15.58 & 45 & 15,\,16 \\
T1 & SDSSp J083717.22-000018.3 & 17.10 & 15 & 15 \\
T2 & SDSSp J125453.90-012247.4 & 14.89 & 29 & 15,\,17 \\
T3 & 2MASS J12095613-1004008 & 15.91 & 15 & 15,\,17 \\
T4 & 2MASSI J2254188+312349 & 15.26 & 27 & 15,\,17 \\
T5 & 2MASS J15031961+2525196 & 13.94 & 19 & 15,\,17 \\
T6 & SDSSp J162414.37+002915.6 & 15.49 & 11 & 15,\,18 \\
T7 & 2MASSI J0727182+171001 & 15.60 & 6 & 15,\,18 \\
T8 & 2MASSI J0415195-093506 & 15.70 & 7 & 15,\,17 \\
T9 & UGPS J072227.51-054031.2 & 16.49 & 2 & 19,\,20 \\
d/sdM4 & LSPM J0713+2151 & 14.26 & 123 & 40 \\
d/sdM5 & 2MASS J2059203+175223 & 15.87 & 39 & 36 \\
d/sdM6 & LSR 1610-0040 & 12.91 & 202 & 37 \\
d/sdM7 & NLTT 57956 & 13.61 & 141 & 1 \\
d/sdM8 & 2MASS J15561873+1300527 & 15.91 & 26 & 38 \\
d/sdM9 & SSSPM 1444-2019 & 12.55 & 178 & 39 \\
d/sdL0 & 2MASS J00412179+3547133 & 15.94 & 31 & 38 \\
d/sdL1 & 2MASS J17561080+2815238 & 14.71 & 42 & 1 \\
d/sdL7 & 2MASS J11582077+0435014 & 15.61 & 37 & 1 \\
sdM2 & LHS 3181 & 13.83 & 173 & 1,\,21 \\
sdM4 & LSPM J0949+1746 & 15.26 & 68 & 40 \\
sdM5 & LHS 407 & 12.82 & 202 & 1,\,21 \\
sdM6 & LHS 1074 & 14.68 & 63 & 1,\,2 \\
sdM7 & LHS 377 & 13.19 & 153 & 1,\,22 \\
sdM8 & 2MASS J01423153+0523285 & 15.91 & 23 & 17 \\
sdM9.5 & SSSPM 1013-1356 & 14.62 & 63 & 22 \\
sdL0 & WISE J04592121+1540592 & 14.96 & 21 & 23 \\
sdL3.5 & SDSS J125637.16-022452.2 & 16.10 & 19 & 24 \\
sdL4 & 2MASS J16262034+3925190 & 14.44 & 53 & 22 \\
esdM0 & LHS 217 & 12.20 & 203 & 1 \\
esdM4 & LHS 375 & 12.15 & 190 & 21 \\
esdM5 & LP 589-7 & 14.50 & 130 & 26 \\
esdM6.5 & LHS 2023 & 14.91 & 78 & 21,\,25 \\
esdM7.5 & APMPM 0559-2903 & 14.89 & 61 & 26 \\
esdM8.5 & LEHPM 2-59 & 15.52 & 34 & 26 \\
M6$\gamma$ & TWA 8B & 9.84 & 81 & 31 \\
M7$\gamma$ & 2MASS J03350208+2342356 & 12.25 & 144 & 31,\,33 \\
M8$\gamma$ & 2MASSW J1207334-393254A & 13.00 & 167 & 32,\,31\\
M9$\gamma$ & TWA 26 & 12.69 & 141 & 32,\,31 \\
L0$\gamma$ & 2MASS J01415823-4633574 & 14.83 & 88 & 31,\,35 \\
L1$\gamma$ & 2MASS J05184616-2756457 & 15.26 & 48 & 31,\,33 \\
L2$\gamma$ & 2MASS J05361998-1920396 & 15.77 & 48 & 31,\,33 \\
L3$\gamma$ & 2MASSW J2208136+292121 & 15.80 & 39 & 14,\,31 \\
L4$\gamma$ & 2MASS J05012406-0010452 & 14.98 & 83 & 14,\,27,\,33 \\
L6$\gamma$ & 2MASSW J2244316+204343 & 16.48 & 58 & 14,\,31,\,34 \\
M8$\beta$ & 2MASSI J0019262+461407 & 12.60 & 228 & 2,\,28,\,33 \\
L0$\beta$ & 2MASSW J1552591+294849 & 13.48 & 147 & 2,\,28,\,33 \\
L1$\beta$ & 2MASS J02271036-1624479 & 13.57 & 104 & 27,\,29,\,30 \\
L2$\beta$ & LSR 0602+3910 & 12.30 & 237 & 2,\,31 \\
L3$\beta$ & 2MASSI J1726000+153819 & 15.67 & 58 & 2,\,14,\,31 \\
\enddata
\tablerefs{
(1) \citet{2010ApJS..190..100K}; 
(2) \citet{2014ApJ...794..143B}; 
(3) \citet{2016ApJ...819...87M}; 
(4) \citet{2003ApJ...596..561M}; 
(5) \citet{2005ApJ...623.1115C}; 
(6) \citet{2006AJ....131.1007B}; 
(7) \citet{knapp:2004:3553}; 
(8) \citet{2007ApJ...658..557B}; 
(9) \citet{2008arXiv0811.0556S}; 
(10) \citet{2007ApJ...659..655B}; 
(11) \citet{chiu:2006:2722}; 
(12) \citet{2006AJ....132..891R}; 
(13) \citet{2006ApJ...639.1114R}; 
(14) \citet{2015ApJS..219...33G}; 
(15) \citet{burgasser:2006:1067}; 
(16) \citet{2007AJ....134.1162L}; 
(17) \citet{burgasser:2004:2856}; 
(18) \citet{burgasser:2006:1095}; 
(19) \citet{cushing:2011:50}; 
(20) \citet{2010MNRAS.408L..56L}; 
(21) \citet{greco:2019:182}; 
(22) \citet{burgasser:2004:l73}; 
(23) \citet{kirkpatrick:2014:122}; 
(24) \citet{burgasser:2009:148}; 
(25) \citet{2005PASP..117..676R}; 
(26) \citet{burgasser:2006:1485}; 
(27) \citet{2018AJ....155...34C}; 
(28) \citet{2016ApJ...833...96L}; 
(29) \citet{burgasser:2008:579}; 
(30) \citet{2008AJ....136.1290R}; 
(31) \citet{2013ApJ...772...79A}; 
(32) \citet{looper:2007:l97}; 
(33) \citet{allers:2010:561}; 
(34) \citet{looper:2008:528}; 
(35) \citet{kirkpatrick:2006:1120};
(36) \citet{2003AJ....126.2487B}; 
(37) \citet{2003AJ....125.1598L}; 
(38) \citet{2004AJ....127.2856B}; 
(39) \citet{Scholz04};
(40) \citet{2005AJ....129.1483L}.
}
\end{deluxetable}

\subsection{Testing and Validation set} 
\label{sec:Testing and Validation set}
To further enrich our data beyond the training sample, we compiled an independent set of 1548 spectra that we used exclusively for validation and testing. Specifically, we pulled 1527 spectra from the Spex Prism Library (SPL; see Table~\ref{tab:testset}) and supplemented them with an additional 21 metal-poor dwarf spectra not in SPL. Of these, 18 unique sources are listed in Table~\ref{table:subdwarf_spectra}, while the remaining 3 are duplicate observations taken at a different time.

We designated 248 of the 1548 \replaced{near-infrared}{NIR} dataset (covering the M, L, and T types) together with 148 synthetic spectra as our validation set to tune model hyperparameters and mitigate overfitting. The synthetic spectra were incorporated to supplement subtypes with insufficient SPL coverage. Details regarding the data generation are provided in Section~\ref{subsec:Building the Testing and Validation Dataset}. 

\begin{figure}[ht]
    \centering
    \includegraphics[width=\columnwidth]{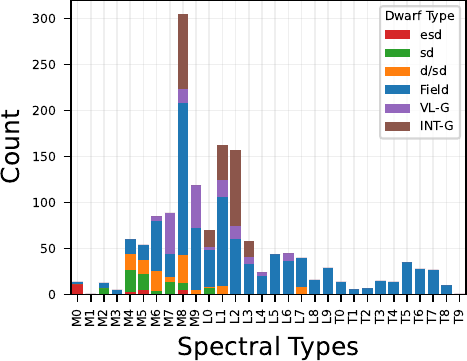}
    \caption{Stacked spectral type distribution of the 1548 spectra. Metallicity and gravity classes are indicated by different color bars. Despite having standards for sdM9.5, sdL3.5, sdL4, and esdM6.5, we do not have examples of these classes in the testing and validation set.}
    \label{fig:Test_set_hist}
\end{figure}

We set aside the remaining 1300 objects as our final testing set. These spectra were not used in any way during training or model tuning. We use them only for the final evaluation of performance metrics. Figure~\ref{fig:Test_set_hist} shows the distribution of spectral types within both the validation and testing sets combined. Because earlier-type M dwarfs typically peak in the red-optical, and well-established classification schemes exist in that region (e.g., \citealt{1997AJ....113..806G}), our near-infrared dataset naturally includes fewer early M types. We also see a lower representation of late L dwarfs and L/T transition objects, both because of their intrinsic faintness and the relative rarity of these subtypes \citep{2003AJ....126.2421C, 2007ApJ...659..655B,2008AJ....135..580R, 2013MNRAS.430.1171D}. 
\begin{deluxetable*}{lcccccccccc}
\tablecaption{SpeX Prism Library Dataset
\label{tab:testset}}
\tablewidth{\textwidth} 
\setlength{\tabcolsep}{3.5pt} 
\tabletypesize{\scriptsize} 
\tablehead{
  \colhead{ShortName} &
  \colhead{Ra} &
  \colhead{Dec} &
  \colhead{SNR} &
  \colhead{SPT$_\mathrm{B06}$} &
  \colhead{SpT$_\mathrm{C18}$} &
  \colhead{SpT$_\mathrm{K10}$} &
  \colhead{J$_\mathrm{2MASS}$} &
  \colhead{H$_\mathrm{2MASS}$} &
  \colhead{K$_\mathrm{2MASS}$} &
  \colhead{Refs.}\\
  \colhead{} &
  \colhead{} &
  \colhead{} &
  \colhead{} &
  \colhead{(Adopted SpT)} &
  \colhead{} &
  \colhead{} &
  \colhead{} &
  \colhead{} &
  \colhead{} &
  \colhead{}
}
\startdata
J0000+2554 & 0.056417  & 25.904999  & 49  & T4          & T4          & T4          & 15.06 & 14.73 & 14.84 & 1 \\
J0000-1245 & 0.119458  & -12.754250 & 165 & M9          & M9          & M9          & 13.20 & 12.45 & 11.97 & 2 \\
J0001+1535 & 0.300750  & 15.593194  & 24  & L3$\gamma$ & L3$\gamma$  & L3$\gamma$  & 15.52 & 14.51 & 13.71 & 3 \\
J0001-0841 & 0.383042  & -8.690556  & 34  & d/sdL1     & d/sdL1     & d/sdL1     & 15.71 & 15.03 & 14.70 & 4 \\
J0001-0943 & 0.476292  & -9.716861  & 49  & M8$\beta$   & M8$\beta$   & M8$\beta$   & 14.79 & 14.17 & 13.82 & 5 \\
\multicolumn{11}{c}{\dots} \\
J2354-1852 & 358.749600& -18.872499 & 118 & L1          & L1          & L1          & 14.18 & 13.44 & 13.04 & 2 \\
J2356-3426 & 359.045010& -34.434555 & 78  & M8$\gamma$ & M9$\gamma$  & M9$\gamma$  & 12.95 & 12.38 & 11.97 & 2 \\
J2356-1553 & 359.228180& -15.886416 & 18  & T5          & T5          & T5          & 15.82 & 15.63 & 15.77 & 1 \\
J2357+1227 & 359.318730& 12.461611  & 8   & T6          & T6          & T6          & 16.52 & 15.75 & 16.12 & 28 \\
J2359-2007 & 359.990110& -20.127611 & 137 & d/sdM8      & d/sdL0      & d/sdL0      & 14.38 & 13.62 & 13.25 & 2 \\
\enddata
\tablecomments{A portion of this table is shown here for guidance regarding its form and content.
The machine--readable version includes all 1527 rows and additional columns--\texttt{Name}, \texttt{Designation}, \texttt{Chisqr}, \texttt{Dwarf\_Type}, \texttt{SpT\_simbad}, \texttt{Parallax}, \texttt{Parallax\_err}, \texttt{Parallax\_ref}, and \texttt{Absmag}--that are omitted from the PDF display.}
\tablerefs{
(1) \citet{burgasser:2006:1067};
(2) \citet{2014ApJ...794..143B};
(3) \citet{knapp:2004:3553};
(4)  \citet{2014ApJ...787..126L};
(5)  \citet{2018AA...619L...8R};
(6)  This study;
(7)  \citet{2006AJ....131.2722C};
(8)  \citet{2016AA...589A..49S};
(9)  \citet{2010ApJS..190..100K};
(10) \citet{2017AJ....154..112K};
(11) \citet{2016ApJ...817..112S};
(12) \citet{2004AJ....127.2856B};
(13) \citet{2008ApJ...689.1295K};
(14) \citet{2015ApJ...814..118B};
(15) \citet{2010ApJ...710.1142B};
(16) \citet{2005AJ....129.1483L};
(17) \citet{2003AJ....126.2421C};
(18) \citet{2007AJ....133..439C};
(19) \citet{2016ApJ...830..144R};
(20) \citet{2022AA...664A.111B};
(21) \citet{2021ApJS..257...45H};
(22) \citet{2011AJ....141...70B};
(23) \citet{2013ApJ...777...84B};
(24) \citet{1999AA...351L...5E};
(25) \citet{2003AJ....125.3302G};
(26) \citet{2016ApJS..224...36K};
(27) \citet{2007ApJ...658..617B};
(28) \citet{2013ApJS..205....6M};
(29) \citet{2017AJ....153..196S};
(30) \citet{2012AJ....144...94G};
(31) \citet{2011ApJS..197...19K};
(32) \citet{2006ApJ...639.1095B};
(33) \citet{2008ApJ...676.1281M};
(34) \citet{2016ApJS..225...10F};
(35) \citet{2009AA...503..639T};
(36) \citet{2009AJ....137....1F};
(37) \citet{2008AJ....136.1290R};
(38) \citet{2008ApJ...681..579B};
(39) \citet{2009ApJ...699..649S};
(40) \citet{2002AJ....123.3409H};
(41) \citet{2003IAUS..211..197W};
(42) \citet{2005PASP..117..676R};
(43) \citet{2009AA...497..619Z};
(44) \citet{2006ApJ...639.1114R};
(45) \citet{2010AJ....139..176F};
(46) \citet{2018AJ....155...34C};
(47) \citet{2014ApJ...792..119D};
(48) \citet{2004AA...418..357K};
(49) \citet{2011ApJ...732...56G};
(50) \citet{2013PASP..125..809T};
(51) \citet{2019ApJ...883..205B};
(52) \citet{2003AA...397..575P};
(53) \citet{2007AJ....133.2320S};
(54) \citet{2014ApJ...783..122K};
(55) \citet{2010ApJ...725L.186D};
(56) \citet{1999ApJ...527..219S};
(57) \citet{2012ApJ...760..152L};
(58) \citet{2007AJ....133.2258S};
(59) \citet{2016ApJ...827...52L};
(60) \citet{2017ApJ...838...73M};
(61) \citet{2010AA...519A..93B};
(62) \citet{1998MNRAS.298L..34F};
(63) \citet{2015ApJS..219...33G};
(64) \citet{2013ApJ...776..126C};
(65) \citet{2014ApJ...784..126E};
(66) \citet{2017AJ....153...46L};
(67) \citet{2007AJ....134..411M};
(68) \citet{2013ApJ...772...79A};
(69) \citet{2006ApJ...645..676L};
(70) \citet{2013AA...557A..43B};
(71) \citet{2004AA...421..763P};
(72) \citet{2007AJ....134.1162L};
(73) \citet{2006AA...460..635C};
(74) \citet{2000AJ....120..447K};
(75) \citet{2008AJ....135.2024A};
(76) \citet{2008AJ....135..785W};
(77) \citet{2012AJ....144...99D};
(78) \citet{2014PASP..126..642S};
(79) \citet{2010AJ....139.1808S};
(80) \citet{2019AJ....157..231K};
(81) \citet{2012ApJ...755...94D};
(82) \citet{2008MNRAS.383..831P};
(83) \citet{2006ApJ...645.1485B};
(84) \citet{2011AJ....141...97W};
(85) \citet{2014AJ....148...91L};
(86) \citet{2006AJ....132.2074M};
(87) \citet{2011AJ....141...71F};
(88) \citet{2010MNRAS.404.1817Z};
(89) \citet{2017AA...598A..92L};
(90) \citet{2008ApJ...674..451B};
(91) \citet{2014MNRAS.437.3603S};
(92) \citet{1979nlcs.book.....L};
(93) \citet{2002ApJ...575..484G};
(94) \citet{2012AA...537A..94S};
(95) \citet{1951AJ.....56Q..34A};
(96) \citet{2009ApJS..184..138H};
(97) \citet{2009AJ....137..304S};
(98) \citet{1952ApJ...116..117V};
(99) \citet{2017ApJS..228...18G};
(100) \citet{2021AA...645A.100S};
(101) \citet{2018ApJ...853...75T};
(102) \citet{2015AJ....150..182K};
(103) \citet{2007AA...468..163D};
(104) \citet{2022ApJ...927..122H};
(105) \citet{2014MNRAS.439.3890G};
(106) \citet{2024AA...685A...6R};
(107) \citet{2011AJ....142...77D};
(108) \citet{2014ApJ...785L..14G};
(109) \citet{2015ApJ...798...73G};
(110) \citet{2007ApJ...657..494B};
(111) \citet{2007ApJ...655..522L};
(112) \citet{2007ApJ...669L..97L};
(113) \citet{2012MNRAS.427.3280F};
(114) \citet{2015AA...582L...5R};
(115) \citet{2005AA...440.1061L};
(116) \citet{2019MNRAS.484.5049M};
(117) \citet{2010AJ....139.2448B};
(118) \citet{2007AJ....133.2825R};
(119) \citet{2014AJ....147...34S};
(120) \citet{2007AJ....133..971A};
(121) \citet{2010AA...515A..92S};
(122) \citet{2007ApJ...659..655B};
(123) \citet{1997AJ....113..806G};
(124) \citet{2007MNRAS.374..445K};
(125) \citet{2014AA...567A...6P};
(126) \citet{2014MNRAS.442.1586D};
(127) \citet{2018AJ....156...76L};
(128) \citet{2020AJ....159..282E};
(129) \citet{2012ApJS..201...19D};
(130) \citet{2012ApJ...757..100D};
(131) \citet{2012AJ....144...62A};
(132) \citet{2003AJ....126.2487B};
(133) \citet{2016AJ....151...46A};
(134) \citet{2014AJ....147...20N};
(135) \citet{2011AJ....142..171G};
(136) \citet{2008ApJ...686..528L};
(137) \citet{2003AJ....125.1598L};
(138) \citet{2004AA...425..519S};
(139) \citet{2018MNRAS.479.2351W};
(140) \citet{2013MNRAS.434.1005Z};
(141) \citet{2010AA...517A..53M};
(142) \citet{2013AJ....146..161M};
(143) \citet{2007ApJ...654..570L};
(144) \citet{2004ApJ...614L..73B};
(145) \citet{2014ApJ...792L..17G};
(146) \citet{2006AJ....131.1007B};
(147) \citet{2015AJ....149..158S};
(148) \citet{2001AJ....121.3235K};
(149) \citet{2004AJ....128..426W};
(150) \citet{2006MNRAS.368.1281P};
(151) \citet{2016Natur.533..221G};
(152) \citet{2000AJ....120.1085G};
(153) \citet{2022ApJ...934..178A};
(154) \citet{2012ApJ...753..142B}.
}
\end{deluxetable*}

\begin{deluxetable*}{lccc}
\tablecaption{Additional subdwarf data not currently in SPL\label{table:subdwarf_spectra}}
\tablehead{
\colhead{Name} & \colhead{Designation} & \colhead{Literature SpT} & \colhead{Data References}
}
\startdata
WISE J0014$-$0838 & J00145014$-$0838231 & sdM9 & 1 \\
2MASS J0115+3130 & J01151621+3130061 & d/sdM8 & 3 \\
WISE J0306$-$0330 & J03060166$-$0330590 & sdL0 & 2 \\
LSPM J0330$+$3504B & J03301720$+$3505001 & d/sdM7 & 4 \\
LEHPM 1-3365 & J03303847$-$2348463 & esdM7 & 4 \\
LSPM J0402+1730 & J04024315+1730136 & sdM7 & 4 \\
HD 114762B & J13121941+1731039 & d/sdM9 & 5 \\
WISE J0435+2115 & J04353580+2115092 &  sdM9 & 2 \\
2MASS J0447-1946 & J04470652-1946392 & sdM7.5 & 6 \\
2MASS J1359+3031 & J13593574+3031039 & d/sdM7 & 3 \\
2MASS J1541+5425 & J15412408+5425598 & d/sdM7 & 3 \\
2MASS J1559$-$0356 & J15590462$-$0356280 & d/sdM8 & 3 \\
2MASS J1640+1231 & J16403197+1231068 & d/sdM9 & 3 \\
2MASS J1640+2922 & J16403561+2922225 & d/sdM7 & 3 \\
LSR 1826+3014 & J18261131+3014201 & d/sdM8.5 & 4 \\
LSR 2036+5059 & J20362186+5059503 & sdM7.5 & 3 \\
WISE J2040+6959 & J20402724+6959237 & sdM9 & 2 \\
LSPM J2331+4607N & J23311807+4607310 & d/sdM7 & 4 \\
\enddata
\tablerefs{1.\cite{2016ApJS..224...36K}; 2.\cite{2014ApJ...781....4L}; 3.\cite{2004AJ....127.2856B}; 4.\cite{2014ApJ...794..143B}; 5.\cite{2009ApJ...706.1114B}; 6.\cite{2010ApJS..190..100K}.}
\end{deluxetable*}

\section{Methods} \label{sec:Methods}

\begin{figure*}[htbp]
    \centering
    \includegraphics[width=\textwidth, trim=4cm 0cm 5cm 0cm, clip]{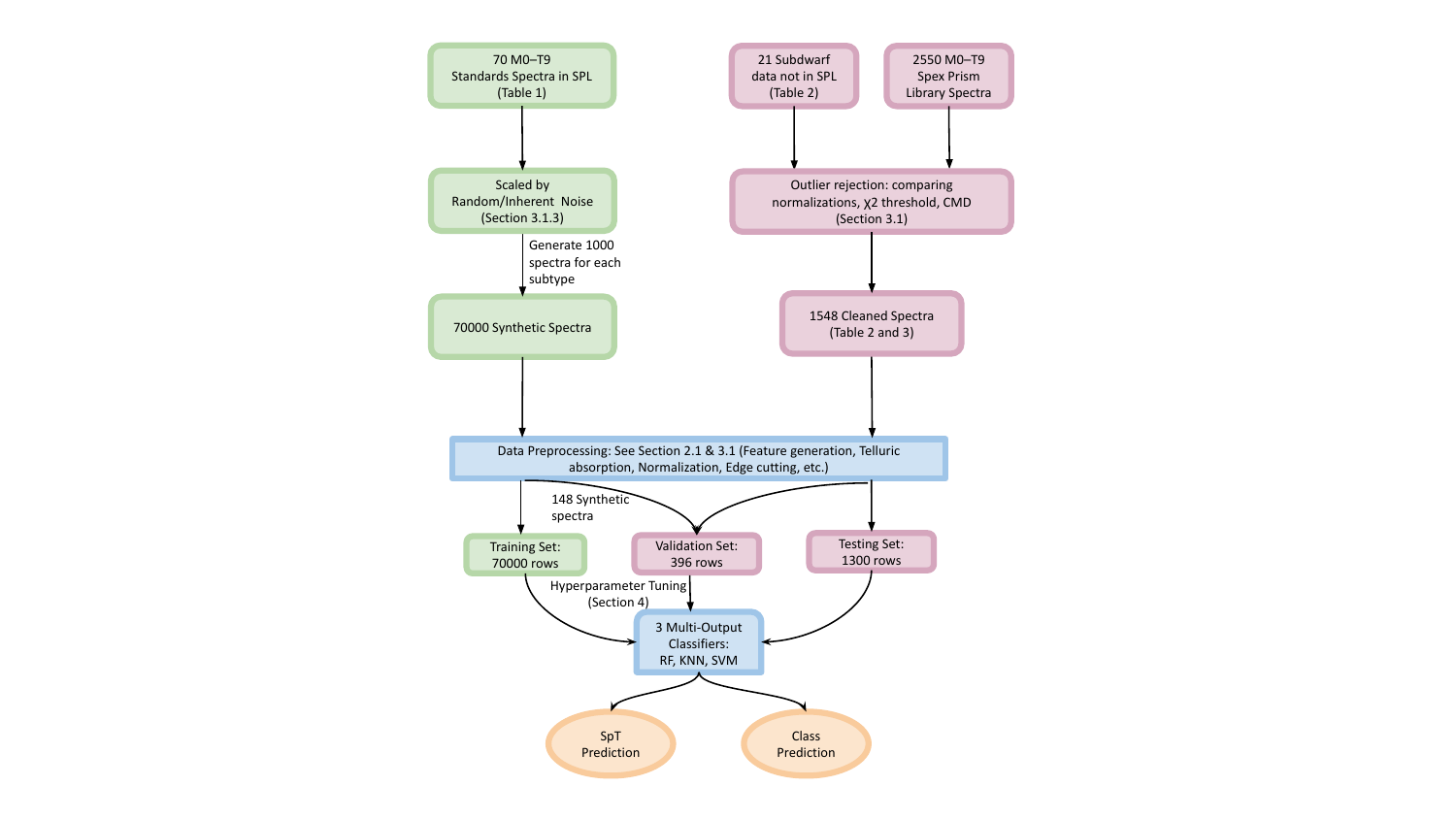}
    \caption{Conceptual workflow diagram for the ML models, detailing data gathering, preprocessing, training, validation, testing, and prediction stages, along with spectral sources and methods.}
    \label{fig:Methodology_diagram}
\end{figure*}

We have created a layered methodology for selecting the data, (pre-)processing the data, training the model, and validating the model. The workflow diagram for our methodology is provided in Figure~\ref{fig:Methodology_diagram}. In this section, we review 
our procedures for data pre-processing, and the construction and tuning of our ML algorithms.

\subsection{Data Preprocessing}
\label{subsec:Data Pre-processing}
The quality of the spectral type labels significantly impacts our ability to infer accurate spectral types for 
unseen data. 
Therefore, we undertook the following steps to ensure data were cleaned and \textit{uniformly} classified before training our models.


\begin{itemize}
    \item All data were constrained to a fit range of [0.87, 2.39]\,$\mu$m to avoid low signal-to-noise data at shorter and longer wavelengths, which is consistent with the normalization range described by C18. 
    \item Spectra were initially normalized to their median fluxes within the \(J\)-band peak between 1.27--1.28~$\mu$m. This range was chosen to sample a pseudo-continuum region common across all spectral types in the sample. 
    This normalization step helps to mitigate potential biases in the ML model arising from variations in flux scaling across the dataset. Further exploration of normalization variations is described in Section~\ref{subsubsec:Normalizations}.    
    \item To prevent contamination by \added{Earth's} atmospheric molecules, we masked the telluric regions at 1.35--1.42 $\mu$m and 1.80--1.95~$\mu$m.
\end{itemize}

\begin{figure}[ht]
    \centering
    \includegraphics[width=\columnwidth]{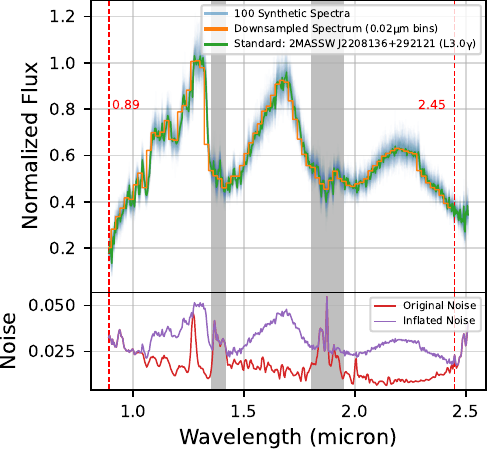}
    \caption{Spectrum of the L3$\gamma$ spectral standard 2MASSW J2208136+292121 (green), along with 100 synthetic spectra in blue drawn from the spectrum and uncertainty. The orange line shows the downsampled 0.02\,\micron\ bins used as features and discussed in Section \ref{subsec:Feature Selection}. 
    We masked out the grey telluric regions at $\sim$1.35 and $\sim$1.9\,$\mu$m. We removed data at the edges, bluer than 0.89\,$\mu$m and redder than 2.45\,$\mu$m as indicated by the red dashed line. In generating the synthetic spectra, we applied the original noise estimates near the spectral edges, while increasing the noise level at intermediate wavelengths to better reflect the uncertainties there. All flux and noises shown are normalized at [1.27,1.28] micron.}
    \label{fig:sample_spectra_combined.pdf}
\end{figure}

These steps were applied to all of the spectral data.
Figure \ref{fig:sample_spectra_combined.pdf} displays an example of the preprocessing procedure applied to the L3$\gamma$ standard 2MASSW J2208136+292121.

\subsubsection{Spectral Classification Labels} 
\label{subsubsec:labels}

The primary labels for our sample are the spectral classifications based on the NIR data. To ensure a common set of labels, we followed the methods described in \citet{2008ApJ...678.1372C} and reclassified all of the sources by minimizing 
the chi-square residual to  our spectral standards,
\begin{equation}
    \chi^2 = \sum \left[ \frac{O_i - \alpha E_i}{\sigma_i} \right]^2,
\end{equation}
where $O_i$ is the source flux at pixel $i$, $\sigma_i$ is the flux uncertainty, $E_i$ is the standard flux, and $\alpha$ is a scale factor to optimize (minimize) the $\chi^2$ value, defined as 
\begin{equation}
    \alpha = \frac{\sum O_i E_i / \sigma_i^2}{\sum E_i^2/ \sigma_i^2}.
\end{equation}


\subsubsection{Varying Normalizations} 
\label{subsubsec:Normalizations}

To ensure the robustness of our preprocessing steps, \deleted{in particular spectral normalization,} \deleted{on classifications, we fit spectra} \deleted{to standard using two common} \deleted{normalization methods described by} \deleted{\citet{burgasser:2006:1067} (B06 in short)} \added{particularly spectral normalization for classifications, we fit spectra to standards using two common normalization methods described by \citet[][hereafter B06]{burgasser:2006:1067}} and C18. The two papers provide two normalization methods: 1. Normalizing the entire NIR band into one section, standardizing the continuum for direct comparison with a reference standard. 2. Normalizing the NIR band in three separate sections:

\begin{enumerate}
    \item 0.87–1.39 $\mu$m for the $zJ$-band,
    \item 1.41–1.89 $\mu$m for the $H$-band, and
    \item 1.91–2.39 $\mu$m for the $K$-band.
\end{enumerate}
By applying separate normalization factors to each sub-band, this approach can better account for localized variations such as atmospheric effects. Note that this method is described in both B06 and C18 with a slightly different wavelength range. In our study, we follow the wavelength regions described in C18, therefore calling method 2 as C18.
For both normalization schemes, we assessed goodness-of-fit by computing the $\chi^2$ statistic over the entire wavelength range used. 

We evaluated the percentage discrepancy and correlation coefficient between the spectral types derived using these normalization choices. Our analysis finds a high correlation coefficient ($R^2$ = 0.96), indicating a strong linear relationship between the two methods, which is not unexpected as both use the same spectral standards for classification. 
The standard deviation of the discrepancies was 1.21, suggesting that while most of the data points are closely aligned, there are some outliers, and provides support to our initial use of an inherent $\pm$1 subtype uncertainty (Section~\ref{subsec:Spectral Standards}). To enhance the robustness of our spectral type classifications, we removed outliers that fall beyond 1.5 standard deviations ($\pm$3 subtype difference) from the mean discrepancy, resulting in a ``clean'' testing and validation sample. This cut corresponds to approximately 3\% of the data, ensuring that the majority of the dataset remains consistent regardless of the choice of normalization within an adopted systematic uncertainty of $\pm$ 3 subtypes. The outliers we removed are listed in Table~\ref{tab:Outliers}.

\begin{deluxetable}{lccccc}
\tabletypesize{\scriptsize}
\tablecaption{Outliers removed from the sample\label{tab:Outliers}}
\tablehead{\colhead{Designation} & \colhead{SpT$_\mathrm{C18}$} & \colhead{SpT$_\mathrm{B06}$} & \colhead{$\Delta$SpT} & \colhead{SNR} & \colhead{Data Ref.} } 
\startdata
J005910.62$-$003850.6 & esdM0 & esdM4 & 4 & 30 & 1 \\
J010024.70+171127.7 & M1 & esdM6.5 & 5 & 108 & 2 \\
J012621.10+142805.7 & L2$\gamma$ & L6$\gamma$ & 4 & 25 & 3 \\
J013525.31+020523.2 & L8 & L4 & 4 & 41 & 4 \\
J013923.88$-$184502.9 & L2 & L6 & 4 & 36 & 4 \\
J020229.29+230513.9 & L6$\beta$ & L2$\gamma$ & 5 & 24 & 5 \\
J032642.25$-$210205.7 & L6$\gamma$ & L2$\gamma$ & 4 & 41 & 6 \\
J032740.95$-$314815.6 & L2 & L6 & 4 & 35 & 7 \\
J032817.43+003257.2 & L2 & L6 & 4 & 82 & 8 \\
J035304.19+041819.3 & L2$\gamma$ & L6$\gamma$ & 4 & 31 & 4 \\
J040418.01+412735.6 & L2 & L6 & 4 & 108 & 9 \\
J040707.52+154645.7 & L2 & L6 & 4 & 96 & 10 \\
J041232.77+104408.3 & L3$\gamma$ & L6$\beta$ & 4 & 12 & 5 \\
J041751.43$-$183832.0 & L2 & L6 & 4 & 18 & 11 \\
J042346.52+084321.1 & L6 & L1$\beta$ & 5 & 15 & 4 \\
J061952.60$-$290359.2 & M9 & L4$\gamma$ & 5 & 36 & 12 \\
J065958.49+171716.2 & L1$\beta$ & L6 & 5 & 71 & 13 \\
J073405.02+581048.4 & esdM0 & esdM4 & 4 & 44 & 2 \\
J082348.18+242857.7 & L2 & L6 & 4 & 85 & 14 \\
J093250.53+183648.5 & L2 & L6 & 4 & 29 & 4 \\
J102921.65+162652.6 & L2 & L6 & 4 & 107 & 8 \\
J102939.58+571544.5 & d/sdL7 & L3 & 4 & 23 & 15 \\
J110604.59$-$190702.5 & L1$\gamma$ & L6 & 5 & 8 & 4 \\
J112208.55+034319.3 & L2 & L6 & 4 & 29 & 4 \\
J113038.04+234148.0 & L2 & d/sdL7 & 5 & 18 & 11 \\
J120703.74$-$315129.8 & L2 & L6 & 4 & 50 & 16 \\
J123526.75+412431.0 & L1$\beta$ & L6 & 5 & 23 & 4 \\
J130428.86$-$003241.0 & L2 & L6 & 4 & 19 & 4 \\
J134514.17+475723.1 & L2 & L6 & 4 & 20 & 4 \\
J141118.48+294851.6 & d/sdL7 & d/sdL1 & 6 & 24 & 11 \\
J141659.87+500625.8 & d/sdL7 & L3 & 4 & 8 & 17 \\
J142227.20+221557.5 & L5 & T1 & 6 & 26 & 17 \\
J145642.68+645009.7 & L6$\gamma$ & L2$\gamma$ & 4 & 32 & 5 \\
J155001.91+450045.1 & L2 & L6 & 4 & 34 & 4 \\
J162255.33+115923.8 & L6 & L2$\beta$ & 4 & 8 & 17 \\
J170726.91+545112.1 & d/sdL1 & d/sdL7 & 6 & 36 & 18 \\
J171531.11+105410.8 & L2 & L6 & 4 & 9 & 4 \\
J173332.50+314458.3 & L2 & L6 & 4 & 74 & 19 \\
J174341.48+212706.9 & L1$\beta$ & L6 & 5 & 51 & 8 \\
J190049.80$-$301650.0 & sdM8 & M4 & 4 & 35 & This study \\
J190049.31$-$414917.6 & esdM0 & esdM8.5 & 8 & 21 & This study \\
J204543.03$-$141131.8 & esdM0 & esdM6.5 & 6 & 48 & 15 \\
J213154.44$-$011937.4 & d/sdL7 & T1 & 4 & 17 & 17 \\
J214205.80$-$310116.2 & L2 & L6 & 4 & 56 & 14 \\
J222736.87$-$185453.1 & L2$\gamma$ & L6$\gamma$ & 4 & 26 & 20 \\
J224253.17+254257.3 & L2 & L6 & 4 & 129 & 14 \\
J224931.09$-$162759.4 & L5 & T0 & 5 & 41 & 20 \\
J225016.27+080824.8 & d/sdL1 & d/sdL7 & 6 & 61 & 11 \\
J233925.27+350716.5 & L2 & L6 & 4 & 51 & 14 \\
\hline
\enddata
\tablerefs{
1. \citet{2011AJ....141...97W},
2. \citet{2010ApJS..190..100K},
3. \citet{2008ApJ...676.1281M},
4. \citet{2017AJ....154..112K},
5. \citet{2017AJ....153..196S},
6. \citet{2015ApJS..219...33G},
7. \citet{2009AJ....137....1F},
8. \citet{2014ApJ...794..143B},
9. \citet{2013ApJ...776..126C},
10. \citet{2008ApJ...681..579B},
11. \citet{2016ApJ...830..144R},
12. \citet{2013ApJ...772...79A},
13. \citet{2019ApJ...883..205B},
14. \citet{2010ApJ...710.1142B},
15. \citet{2016ApJS..224...36K},
16. \citet{2007AJ....133.2320S},
17. \citet{2006AJ....131.2722C},
18. \citet{2016ApJ...817..112S},
19. \citet{2013PASP..125..809T},
20. \citet{2015ApJ...814..118B}
}
\end{deluxetable}

%
We investigated the outliers of the comparison process by visual inspection and found that most outliers are L dwarfs and esdMs. Specifically, for esdMs, we found that the normalization range described by C18---[0.87, 1.39], [1.41, 1.89], and [1.91, 2.39]\,$\mu$m---struggles to find the optimal best-fit as the flux peak of esdMs typically occurs shortward of 0.87\,$\mu$m.  This is expected as the C18 methodology was based on L dwarfs.

\subsubsection{Color Magnitude diagram} 
\label{subsec:Color Magnitude diagram}
The Color-Magnitude Diagram (CMD) is a powerful tool for distinguishing different stellar populations based on their luminosity and color. For cool dwarfs, which include M, L, and T types, there is a well-defined sequence where \replaced{later}{M and L} spectral types \replaced{are}{get} redder and fainter\added{, and T dwarfs get bluer.} \citep{2002AJ....124.1170D, 2005ARA&A..43..195K}. Deviations from this sequence can indicate potential outliers such as giants, which are more luminous for a given color \citep{2018A&A...616A..10G}, unresolved binaries, which appear brighter due to combined light \citep{2018A&A...616A..10G, 2020ApJ...901...49L}, and young stellar objects (YSOs) with circumstellar disks, which may exhibit infrared excesses affecting their colors \citep{1984ApJ...287..610L, 2010ApJS..186..111L}. 

\begin{figure}[ht]
    \centering
    \includegraphics[width=\columnwidth]{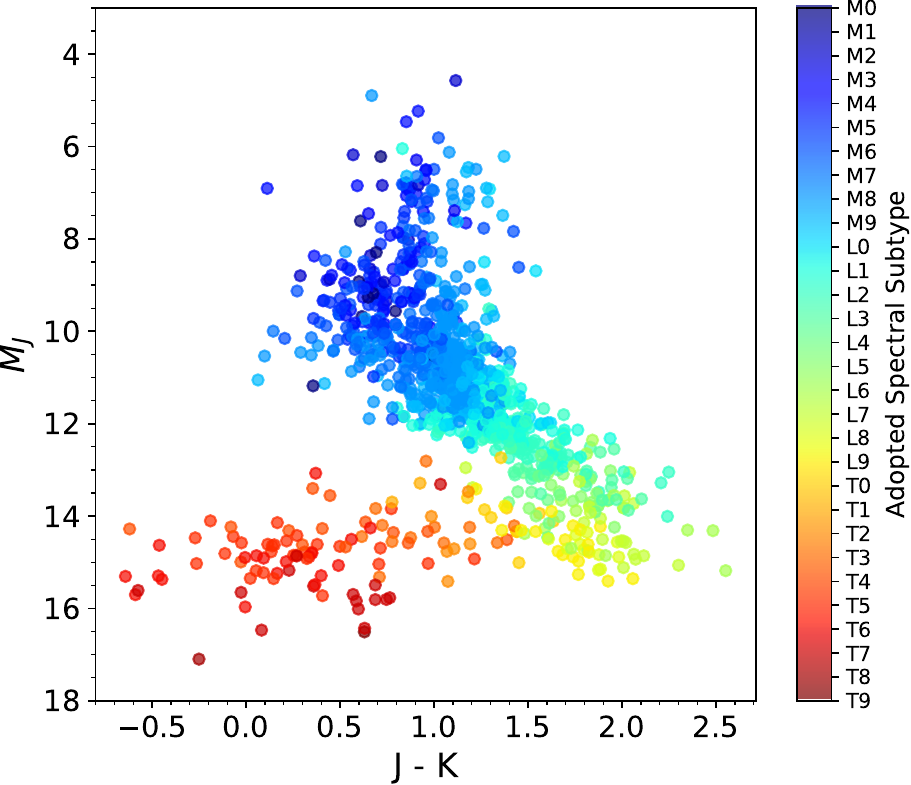}
    \caption{Color-Magnitude Diagram showing absolute \textit{J} magnitude ($M_J$) versus \textit{J-K} color (after outliers removal) for SpeX Prism Library objects with adopted spectral types from M0 to T9. This plot confirms the consistency of our adopted spectral typing method. Objects that deviate from the sequence, including giants, unresolved binaries, and YSOs, are identified and removed to ensure a homogeneous sample.}
    \label{fig:CMD}
\end{figure}

We aim to exclude these objects from our study as they could not be accurately represented by the standards listed in Table~\ref{tbl:standards}, and thus will be outside the parameter space of our models. We used Gaia Data Release 3 parallaxes (\citealt{2021AA...649A...6G}) and 2MASS \textit{JHK} photometry (\citealt{2006AJ....131.1163S}) to obtain $J-K$ and \deleted{absolute} $M_J$ \deleted{magnitudes} for our sources. The adopted spectral types are obtained by methods described in Section~\ref{subsubsec:labels} and \ref{subsubsec:Normalizations}. SPL objects with available parallaxes and \textit{JHK} magnitudes are plotted in Figure~\ref{fig:CMD}. With the help of the CMD, we verified the correctness of our adopted spectral types. We also visually inspected 43 outliers, including YSOs, giants, binaries, and erroneous data not labeled in SPL. This results in a dataset of 1,548 objects for model validation and testing.
\label{sec:CMD}




\subsubsection{Building the Training Set: Synthetic Spectra}
\label{subsubsec:Building the Training set}

\begin{figure}[htbp]
  \centering
  \subfloat[Noise scale vs.\ accuracy\label{fig:noisescale}]{
    \includegraphics[width=0.46\textwidth]{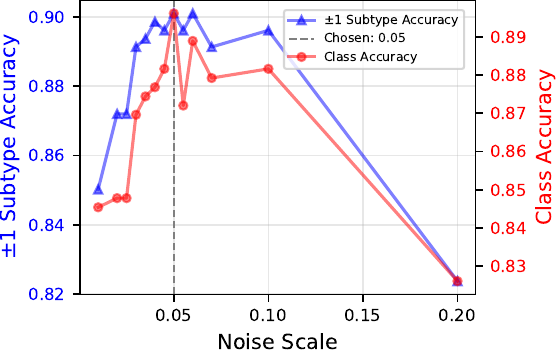}
  }
  \hfill
  \subfloat[Training time vs.\ synthetic data size\label{fig:traintime}]{
    \includegraphics[width=0.46\textwidth]{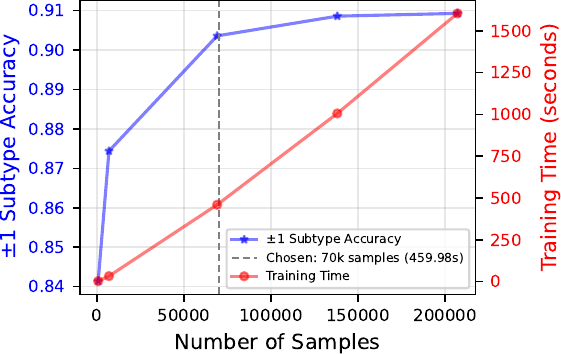}
  }
  \caption{Random Forest metrics for synthetic spectra optimization.
    \textit{Top panel}: The relationship between noise scaling (fractional uncertainty on the flux) and classification accuracy (subtype and class), indicating an optimal noise scale of $5\%$.
    \textit{Bottom panel}: The relationship between the number of synthetic training spectra, training time, and classification accuracy\added{, which shows little improvement in accuracy but increasingly longer training times beyond 70,000 samples}.
  }
  \label{fig:combined}
\end{figure}

\deleted{The training of machine learning} \deleted{approaches requires a large, diverse training} \deleted{set that exhibits sufficient variance to sample} \deleted{the full range of features.}
\deleted{We expanded our initial standard sample} \deleted{of 70 spectra, hereafter ``intrinsic" spectra,} \deleted{by supplementing it with ``synthetic" spectra.}
\deleted{These spectra were generated by randomizing} \deleted{the flux at each pixel assuming} \deleted{a normal distribution centered on the} \deleted{measured value and a standard deviation} \deleted{set by the flux uncertainty,} \deleted{with a minimum standard deviation} \deleted{of 5\% (SNR $= 20$).}
\deleted{This floor on the standard deviation} \deleted{was informed by testing various} \deleted{scaling values on the noise}
\deleted{from 1\% to} \deleted{20\% and choosing} \deleted{the set with the optimal subtype} \deleted{and class uncertainty (Figure~\ref{fig:noisescale}).} 

\added{The training of our machine learning models requires a large, diverse training set. We expanded our initial sample of 70 spectral standards by generating 1,000 synthetic spectra from each one. These synthetic spectra were created by randomizing the flux at each wavelength bin. For each point in a standard's spectrum, we drew a new flux value from a Gaussian distribution centered on the original flux.

The standard deviation ($\sigma$) for this distribution was carefully chosen to be representative of real-world noise. For each flux point, the $\sigma$ was set to the maximum of either the intrinsic uncertainty of the standard (original noise) or a baseline fractional uncertainty (inflated noise generated using a ``noise scale" parameter) of the flux value. The noise scale parameter was not arbitrarily chosen; we treated it as a hyperparameter and optimized it by evaluating the final model's classification accuracy on our validation set. As shown in Figure~\ref{fig:noisescale}, a noise scale of 5\% yielded the best performance.

This methodology ensures that our synthetic data accurately captures the variations observed within a given spectral type. For low-SNR standards (e.g., late T dwarfs) where the intrinsic noise is high, that high uncertainty is propagated directly into the synthetic spectra. This results in a wider distribution of synthetic flux values, which realistically simulates the large measurement uncertainties inherent in observing faint targets and makes our model more robust to noisy data.}

To determine the number of synthesized samples, we evaluated a combination of classification accuracy and training time as \added{a} function of \added{the} number of synthetic spectra, shown in Figure~\ref{fig:traintime}.
We observe that accuracy improves steadily as we add more synthetic spectra, but beyond approximately 70,000 samples, additional gains become marginal while training time continues to increase. Thus, we select 70,000 as the optimal balance between improved accuracy and acceptable computational cost, corresponding to 1,000 synthetic spectra per standard.

\subsubsection{Building the Testing and Validation Sets} \label{subsec:Building the Testing and Validation Dataset}

Starting with 1548 intrinsic spectra, we randomly split the data into 1300 spectra for testing and 248 spectra for validation. For any spectral subtype with fewer than six intrinsic spectra in the validation set, synthetic data (see Section~\ref{subsubsec:Building the Training set}) was added to reach a minimum of 6 samples per subtype. This augmentation introduced an additional 148 synthetic spectra into the validation set, ensuring robust representation of rare subtypes (e.g., late L/T dwarfs, subdwarfs, early M dwarfs) while avoiding overfitting. In summary, the testing set contains 1300 intrinsic spectra, while the validation set—augmented with synthetic data to ensure a minimum of six samples per subtype—comprises 248 intrinsic plus 148 synthetic spectra (a total of 396), effectively representing \added{a} 75/25 split for \added{the} testing and validation sets.

\begin{figure}[htbp!]{}
    \centering
    \includegraphics[width=\columnwidth]
    {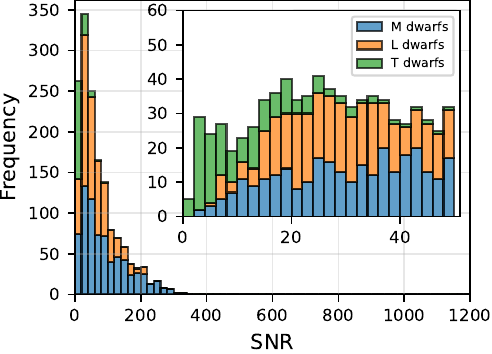}
    \caption{Distribution of SNR for our validation and testing sample. Spectral classes of M, L, and T are distinguished by blue, orange, and green bars, in bins of 20 SNR. 
    The inset plot shows a more granular view of sources with SNR $\leq 50$, in bins of 2 SNR. The majority of very low SNR ($<10$) sources are T dwarfs.}
    \label{fig:snr_hist}
\end{figure}

The distribution of signal-to-noise ratios for the 1548 sample is plotted in Figure~\ref{fig:snr_hist}, with most sources having SNR $\leq$ 100 and a subset of very low SNR (SNR $\leq$ 10) spectra dominated by T dwarfs. 

\subsubsection{Feature Selection} \label{subsec:Feature Selection}

Feature selection impacts the model's ability to recognize relevant patterns in the data. Different features can drastically affect the model's performance and efficiency. In this study, we utilize \textit{average} fluxes within wavelength bins as the features for the models. 

\added{To put these bin sizes in context, the low-resolution SpeX prism data have a resolving power of $R = \lambda/\Delta\lambda \approx 120$. This corresponds to a native spectral resolution that varies across our wavelength range, from $\sim$0.007~$\mu$m at 0.89~$\mu$m to $\sim$0.02~$\mu$m at 2.45~$\mu$m. }We tested the average flux over different bin sizes: 0.01, 0.02, 0.03, 0.04, and 0.10~$\mu$m. We found that using 0.02~$\mu$m provides the best performance. \deleted{, possibly because it is small} \deleted{enough to capture significant spectral} \deleted{variations indicative of different} \deleted{spectral types.} \added{This bin size effectively preserves the native resolution of the data at redder wavelengths while moderately binning the blue end by a factor of ~3, balancing spectral detail with signal-to-noise. } Smaller bin sizes (e.g., 0.01 $\mu$m) could be too sensitive to small fluctuations in the data, capturing more noise than actual spectral variations. On the other hand, larger bin sizes (e.g., 0.10 $\mu$m) might oversmooth the data, losing important spectral features. 
We have included the respective performance of all other bins in Table \ref{table:performance_metrics_of_different_bins} in Appendix section \ref{sec:Appendix_binning}. 

During the preliminary analysis, it was noted that the flux values within the range shortward of 0.89~$\mu$m are missing for several early-type \textsc{vl-g} L dwarfs spectra, including the L3$\gamma$ standard plotted 
in Figure~\ref{fig:sample_spectra_combined.pdf}. Consequently, we refined the range for feature generation, limiting it to 0.89--2.45~$\mu$m to further diminish the impact of very low SNR regions and address the missing value problem \citep{2003PASP..115..362R}. 
In total, we generated 67 features from 0.89--2.45~$\mu$m using the 0.02 $\mu$m bin, excluding the telluric absorption region.

\section{Results}
\label{sec:results}
In this section, we present the performance of the tested models. Each algorithm used in this study is an example of a ``supervised" ML algorithm, i.e., the data is labeled. In our specific case, we are using supervised ML to predict the spectral types (labels) of low-mass stars and brown dwarfs. 
Supervised ML involves training a model on labeled data, enabling it to learn and recognize patterns from the input. When a new, unlabeled input is encountered, the model can determine the best-fit output. This method has proven successful in various ML problems \citep[e.g.,][]{2022MNRAS.513..516F, 2023PASP..135d4502S}.

Here we discuss all three classifiers, RF, SVM, and KNN, in depth. We convert all algorithms to output both spectral type (M0--T9) and class (field, d/sd, sd, esd, \textsc{vl-g}, \textsc{int-g}) in parallel, using the \texttt{MultiOutputClassifier}\footnote{\url{https://scikit-learn.org/stable/modules/generated/sklearn.multioutput.MultiOutputClassifier.html}} implemented in the \texttt{scikit-learn} library. All models output a 2-dimensional array, with the first integer representing spectral type, and the second integer representing classes. 
This method reduces a 70-class classification problem into a 29 SpT-class and a 6 metallicity/gravity-class classification problem, making it significantly easier to visualize and analyze the results. 
To evaluate our model's performance, we use accuracy as the metric of the model performance. Accuracy measures the percentage of true labels (adopted spectral type) out of the total number of classifications. Since our training data contains 1000 synthetic spectra for each subtype, this creates a uniform dataset in which accuracy can be used as a precise metric.
\subsection{Random Forest Classifier} 
\label{subsubsec:RF ClassifierResults}

The Random Forest algorithm \citep{598994} is an ensembling method created from a group of individual decision trees.
For a multi-class classification problem, each decision tree's output counts as a vote, and the label with the highest votes is assigned as the final output for the instance. RF carries out bootstrapping, which randomly samples the training set with replacement to create multiple subsets. Each subset is used to grow a tree, and only a random subset of features is used when splitting the nodes. 
This introduces randomness and diversity, making the model more robust. 
\subsubsection{RF Parameter Tuning} \label{subsubsec:RF Parameter Tuning}
RF has several hyperparameters such as the maximum depth of decision trees (\texttt{max\_depth}), the number of features to consider when looking for the best split (\texttt{max\_features}), the total number of decision trees in the forest (\texttt{n\_estimators}), the minimum number of samples required to split an internal node (\texttt{nmin\_samples\_split}), and the minimum number of samples required to be at a leaf node (\texttt{nmin\_samples\_leaf}). In the list below, we introduce the definition of these parameters and the most optimal value found by applying the model to the validation set.

\begin{enumerate}
    \item Maximum Depth of Decision Trees: Controls the depth of decision trees to prevent overfitting by restricting complexity. We set this parameter to allow trees to have maximum split to capture complex interactions in the data (\texttt{max\_depth = None})
    \item Number of Features for the Best Split: Controls the diversity of the model. The algorithm will only consider the square root of the total number of features available to find the best split, limiting the search to improve generalization. (\texttt{max\_features = sqrt})
    \item Number of Decision Trees: Controls the number of decision trees. A large number of trees ensures that the final prediction is stable and reliable (\texttt{n\_estimators = 700})
    \item Minimum Samples for Splitting Node: Minimum number of samples required to split an internal node. This ensures that splits are only made when there is sufficient data, which helps avoid decisions based on noise.(\texttt{min\_samples\_split = 5})
    \item Minimum Samples at Leaf Node: Minimum number of samples required at a leaf node. This permits maximum splitting to capture complex patterns (\texttt{min\_samples\_leaf = 1})
\end{enumerate}

\subsubsection{RF Feature Importance} \label{subsubsec:RF Gini Feature Importance}

\begin{figure*}[htbp]{}
    \centering
    \includegraphics[width=\textwidth]{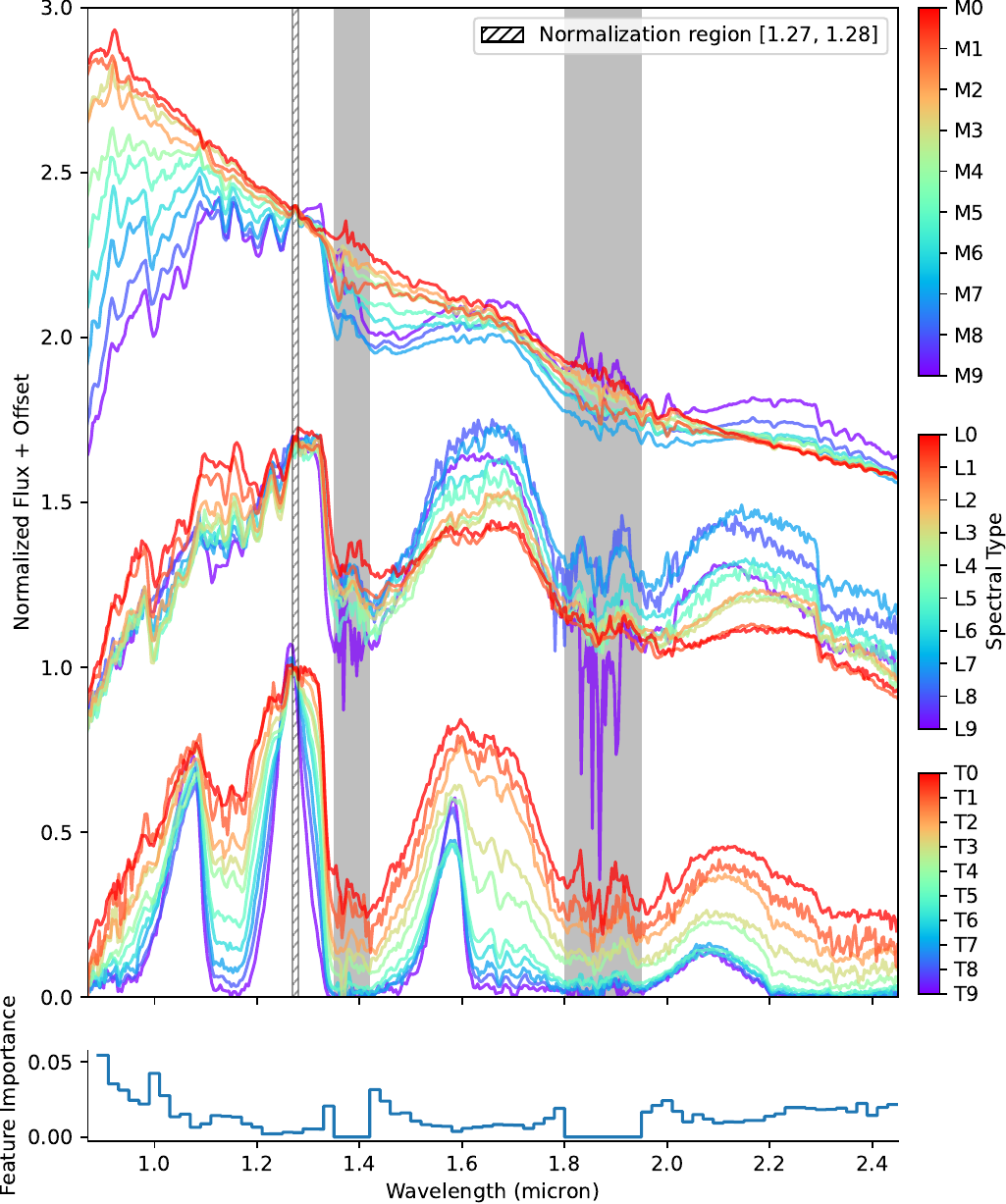}
    \caption{Top: All M, L, and T field Spectral Standards plotted together in order. All spectra are normalized around 1.27 $\mu$m, and different spectral types are separated by an offset. The shaded region is the telluric absorption that was masked. \added{The normalization region (1.27--1.28\,$\mu$m) used to scale each spectrum is indicated by the pair of solid vertical lines and the hatched band.} Bottom: The relative feature importance score. \replaced{The feature importance is an attribute from Scikit-Learn's}{This feature importance is an attribute specific to the} Random Forest classifier.}
    \label{fig:Combined_Spectra_featureImportance}
\end{figure*}

\begin{figure*}[htbp]{}
    \centering
    \includegraphics[width=\textwidth]{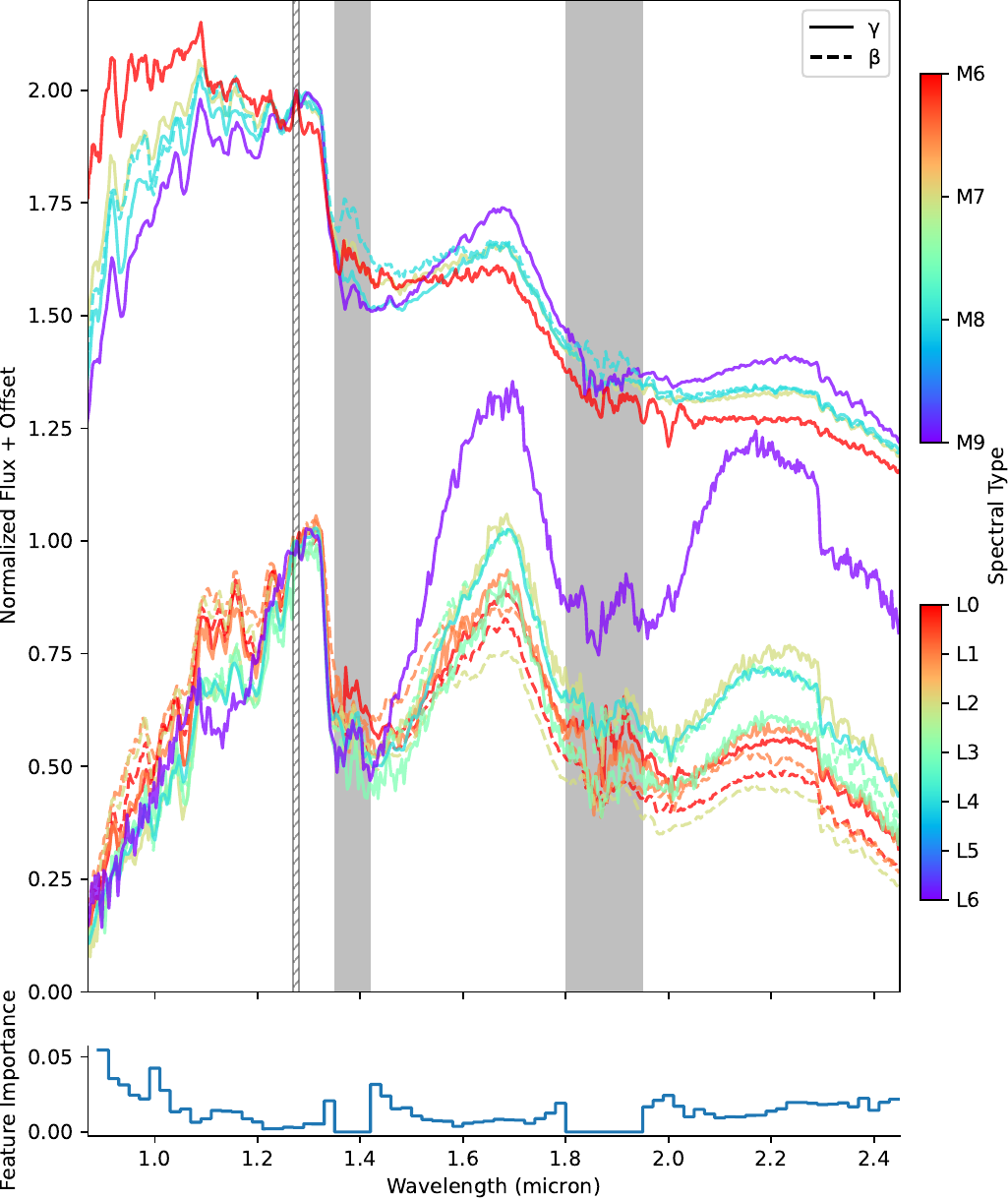}
    \caption{Same as Figure~\ref{fig:Combined_Spectra_featureImportance}, but for low and intermediate surface gravity objects.}
    \label{fig:Gravity_with_Feature_Importances}
\end{figure*}

\begin{figure*}[htbp]{}
    \centering
    \includegraphics[width=\textwidth]{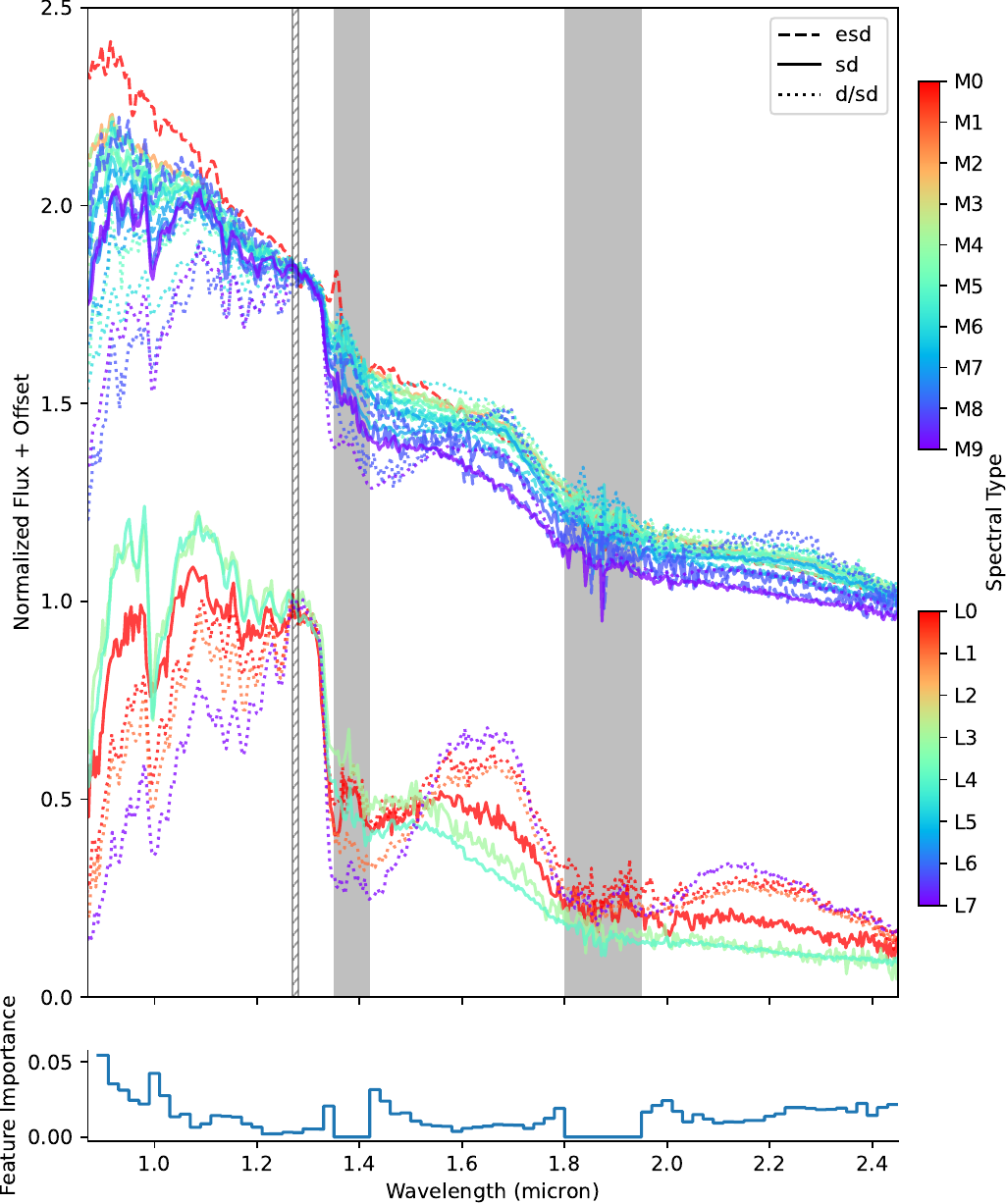}
    \caption{Same as Figure~\ref{fig:Combined_Spectra_featureImportance}, but for metal-poor objects.}
    \label{fig:Subdwarfs_with_Feature_Importances}
\end{figure*}

The RF model, based on decision trees, allows us to calculate the relative feature importance score of each feature. We determined and plotted the relative feature importance score in Figure \ref{fig:Combined_Spectra_featureImportance}, \ref{fig:Gravity_with_Feature_Importances}, and \ref{fig:Subdwarfs_with_Feature_Importances}. All standards used are plotted with RF feature importance across all wavelengths. \deleted{M, L, and T standards were separated by an} \deleted{offset and normalized to the} \deleted{1.27--1.28\,$\mu$m range.}

The feature importance score is calculated as:
\[
FI(f) = \frac{1}{N} \sum_{t=1}^{N} \sum_{n \in M_f} w_n \Delta I(n)
\]
where \( M_f \) represents the set of nodes in each tree that use feature \( f \) for splitting, and 
\[
\Delta I(n) = I(n) - w_{\text{left}} I(\text{left}) - w_{\text{right}} I(\text{right})
\]
with
\[
I_G(p) = 1 - \sum_{i=1}^{c} p_i^2
\]
\begin{itemize}
    \item \( I_G(p) \): Gini impurity. This is the likelihood of misclassifying a randomly chosen sample if it were labeled according to the distribution of labels in the node.
    \item \( p_i \): Proportion of samples belonging to class \( i \)
    \item \( c \): Number of classes
    \item \( FI(f) \): Feature importance of feature \( f \). This quantifies how much the feature f contributes to reducing impurity across the entire forest.
    \item \( N \): Number of trees in the forest
    \item \( M_f \): All nodes that use feature \( f \) for splitting
    \item \( w_n \): Weight of node \( n \), computed as the number of samples reaching node \( n \) divided by the total number of samples
    \item \( \Delta I(n) \): The difference between the impurity of the parent node  and the combined impurities of the child nodes, each weighted by the number of samples
    \item \( I(n) \), \( I(\text{left}) \), \( I(\text{right}) \): Impurity of node \( n \), left child node, and right child node, respectively
\end{itemize}

The feature importance analysis reveals a prominent peak in the 0.89--0.91 $\mu$m range, followed by secondary peaks at 0.99--1.01 $\mu$m and the H$_2$O absorption feature at 1.4 $\mu$m. Overall, the \replaced{$J$-band}{Z and Y band} emerges as the most critical spectral region for classifying these cool objects, \deleted{aligning with established classification}  \deleted{frameworks such as those proposed by} \deleted{\citet{2010ApJS..190..100K}}\added{as they show the largest variance in fluxes.} Additionally, the model assigns significant weight to fluxes near the 1.4 $\mu$m H$_2$O band, indicating its potential importance despite the challenges posed by telluric contamination. A more detailed analysis of feature importance, specifically tailored to each spectral class, is provided in Section \ref{sec:Discussion}. \added{It is critical to note, however, that the feature importance at the J-band normalization window is artificially suppressed by our choice of normalization region (1.27--1.28 $\mu$m). This process minimizes the variance in that part of the spectrum, leading to a near-zero importance score, and makes a direct comparison of the $J$-band's overall importance to the H and K bands challenging.}




\subsubsection{RF Result} 
\label{subsubsec:RF Result}

\begin{figure*}[ht]{}
    \centering
    \includegraphics[width=\textwidth]
    {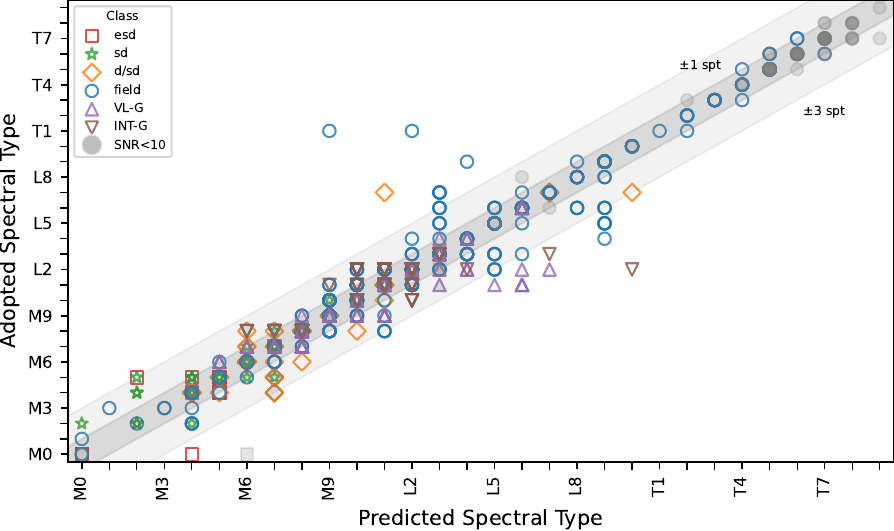}
    \caption{\deleted{Scatter Plot for the} Random Forest (RF) model \replaced{in classifying}{classification of} spectral subtypes. All field and peculiar subtypes are included and grouped under their base type (i.e., L6$\gamma$ goes under L6). This graph demonstrates the predicted label of the RF model \replaced{versus}{compared to} the adopted label defined in the dataset. The dark gray and light gray shaded regions represent $\pm$1 SpT and $\pm$3 SpT, respectively. Lastly, gray markers represent sources with \replaced{extremely}{very} low SNR (SNR $\leq$ \replaced{5}{10})}
    \label{fig:RF_scatter_actual_vs_predicted}
    
\end{figure*}

\begin{figure}[ht]
    \centering
    \includegraphics[width=\columnwidth]{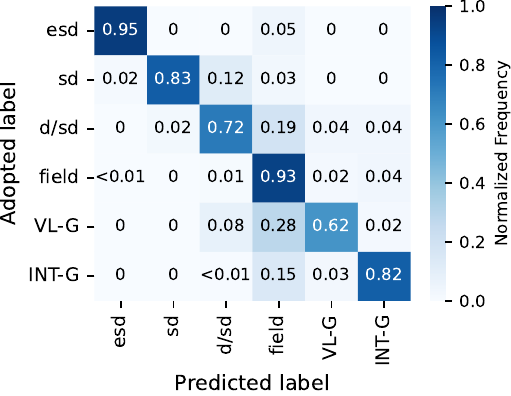}
    \caption{Confusion Matrix for the RF model in classifying classes. All field and peculiar classes are included. The values are normalized across each row (True label), i.e., each row sums to 1. This graph demonstrates the predicted dwarf label by the RF model versus the actual label defined in the dataset. \added{The RF model performs reasonably well for metallicity and \textsc{int-g} classes, but struggles more with the \textsc{vl-g} class.}}
    \label{fig:rf_dwarf_confusion_matrix}
\end{figure}

The multi-output Random Forest (RF) algorithm results in a classification accuracy of $91.5\pm0.8$\% within a SpT range of $\pm$1, and $98.2\pm0.4$\% within a SpT range of $\pm$3, plotted in Figure~\ref{fig:RF_scatter_actual_vs_predicted}. 
In addition to classifying spectral type, the multi-output classifier can classify metallicity class (e.g., subdwarf) and gravity class (e.g., \textsc{int-g}) at $86.2\pm1.0$\% accuracy, shown in Figure \ref{fig:rf_dwarf_confusion_matrix}. A noticeable decline in accuracy is observed for the classification of \textsc{vl-g} L dwarfs, which could be due to atmospheric conditions such as the presence of condensate clouds and turbulent weather patterns exhibiting higher spectral variability and noise levels \citep{2014ApJ...793...75R}. \added{Furthermore, the RF model exhibits a slight systematic bias where it classifies these low-gravity objects as cooler than their adopted types. As this trend is not prominent in the other models (see Figures~\ref{fig:SVM_scatter_actual_vs_predicted} and \ref{fig:KNN_scatter_actual_vs_predicted}), it is more likely an algorithmic artifact of the RF classifier rather than a universal physical effect.} Overall, the model shows consistent results in classifying M and T dwarfs. Complete visualization of the model performance is shown in Appendix \ref{sec:Appendix_classification}, Figure \ref{fig:rf_SpT_confusion_matrix}.


\subsection{Support Vector Machine} \label{subsubsec:SVMResults}

Support Vector Machine (SVM) is an algorithm that classifies data by finding the best hyperplane (a boundary that separates different classes), ensuring the maximum margin between the support vector (closest point to the hyperplane) and the hyperplane. Similar to \added{the} random forest, we implemented the Support Vector Classifier (SVC) algorithm from \texttt{scikit-learn}. Since our problem is a multi-class classification problem, SVC cannot easily find a hyperplane that separates all classes. By default, SVC uses the one-vs-one (OvO) approach for multi-class classification. In OvO, a separate binary classifier is trained for each pair of classes\replaced{.}{, so} $N(N-1)/2$ classifiers are trained for $N$ classes. \deleted{Each classifier is trained to distinguish} \deleted{between one pair of classes.}
During the prediction phase, all 
$N(N-1)/2$ classifiers are consulted, and the class that is selected most frequently is chosen as the final prediction. In \texttt{scikit-learn}'s SVM implementation, this is the default strategy for multi-class classification. 

Another method for multi-class classification in SVM is the one-vs-rest (OvR) strategy. In OvR, a separate binary classifier is trained for each class, treating that class as the positive class and all other classes combined as the negative class. Therefore, if there are 
$N$ classes, $N$ classifiers are trained. During the prediction phase, each classifier predicts a score indicating the confidence that the observation belongs to the respective class. The class with the highest score is chosen as the final prediction. In our study, we opted for the OvO approach because it yielded the highest validation accuracy. Moreover, previous research has demonstrated that OvO requires less training time and is more practical for multi-class classification tasks \citep{Chih2002}.



\subsubsection{SVM Parameter Tuning}
\label{subsubsec:SVM Parameter Tuning}
We followed the same methodology as in Section~\ref{subsubsec:RF Parameter Tuning}. Here are the chosen configurations:

\begin{enumerate}
    \item Regularization Parameter \textsc{(c)}: Controls the trade-off between achieving a low training error and a low testing error, thereby enhancing the model's generalization to unseen data. After parameter tuning, a value of 10 was selected as it provided the best balance between bias and variance, minimizing overfitting while maintaining high accuracy.
    
    \item \textsc{kernel}: Specifies the kernel type to be used in the algorithm, enabling the model to capture complex decision boundaries. The radial basis function (`rbf') kernel was chosen because it consistently yielded superior performance during parameter tuning, effectively handling the non-linear relationships in the spectral data.

    \item \textsc{gamma}: Defines how far the influence of a single training example reaches. Low values mean it reaches far, while high values mean it only considers points close to the decision boundary. We set this value to `scale' (using $[{N \cdot \text{Var}(X)}]^{-1}$) as it automatically adjusts based on the input features, leading to optimal performance across different feature scales observed in our dataset.
 
    \item \textsc{decision\_function\_shape}: Specifies the strategy for multi-class classification. The OvO approach was selected because it demonstrated higher classification accuracy in multi-class scenarios during validation, effectively distinguishing between multiple spectral subtypes. 
\end{enumerate}

The multi-output Support Vector Machine (SVM) achieves a $\pm$1 SpT accuracy of $91.1\pm0.8\%$ and a $\pm$3 SpT accuracy of $98.9\pm0.3\%$. The scatter plot illustrating the classification across spectral types is presented in Figure \ref{fig:SVM_scatter_actual_vs_predicted}, with detailed performance metrics shown in Appendix Section \ref{sec:Appendix_classification}.
Similarly to the RF model, the SVM shows a decline in accuracy for L dwarfs. In addition to classifying spectral type, the classifier classifies metallicity class (e.g., sd) and gravity class (e.g., \textsc{int-g}) at $82.7\pm1.0\%$ accuracy, shown in Figure \ref{fig:rf_dwarf_confusion_matrix}. 

\begin{figure*}[ht]{}
    \centering
    \includegraphics[width=\textwidth]
    {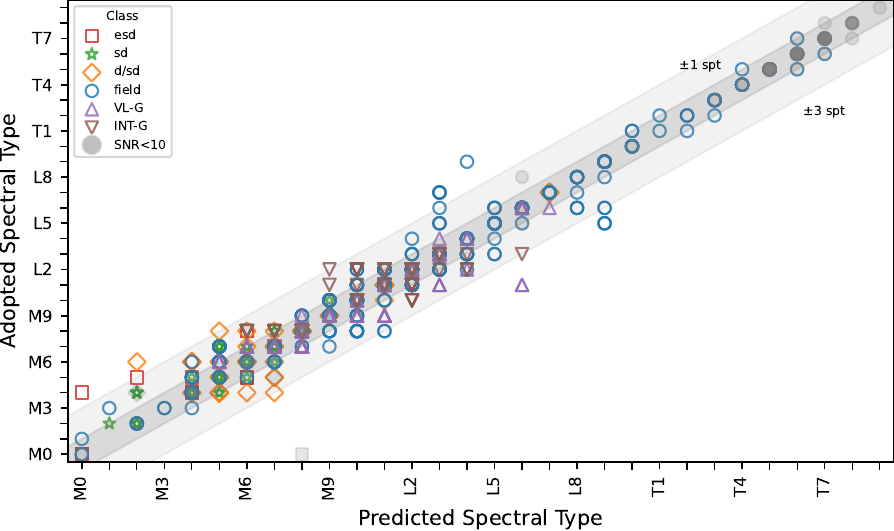}
    \caption{Same as Figure~\ref{fig:RF_scatter_actual_vs_predicted} for the SVM model.}
    \label{fig:SVM_scatter_actual_vs_predicted}
\end{figure*}

\begin{figure}[ht]{}
    \centering
    \includegraphics[width=\columnwidth]
    {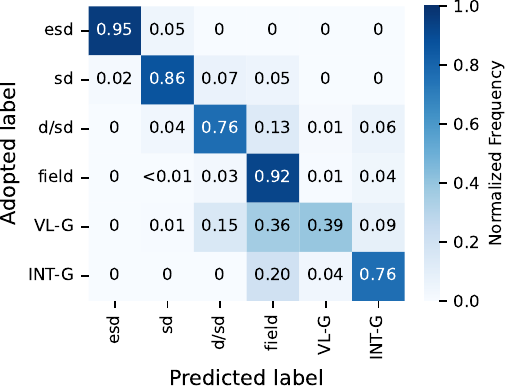}
    \caption{Same as Figure~\ref{fig:rf_dwarf_confusion_matrix} for the SVM model.}
    \label{fig:svm_dwarf_confusion_matrix}
\end{figure}

\subsection{K-Nearest Neighbors} \label{subsec:KNNResults}

K-Nearest Neighbors (KNN) is one of the simplest algorithms in supervised ML. It looks at the closest k-neighbors and classifies the target based on majority votes. As spectra of the same subtype have similar absorption features and morphology, the nearest neighbors method, based on distance metrics, can be extremely useful for detecting similar spectra. In this study, we use the Euclidean distance
to determine the nearest neighbors. The default distance metric of KNN is Minkowski, and with $p=2$. This metric is equivalent to the standard Euclidean metric. 
We tested different values of $k$ from 1 to 200 and found that $k=29$ leads to the best performance, with a decreasing trend in accuracy for increasing $k$ after $k=100$. $k=29$ is relatively large to smooth out the decision boundaries, making it less sensitive to individual instances. A smaller value will make the model more sensitive to local patterns, whereas a larger value will make the model more generalized. We determined $k=29$ was an optimal number to balance between bias and variance.

\begin{figure*}[ht]{}
    \centering
    \includegraphics[width=\textwidth]
    {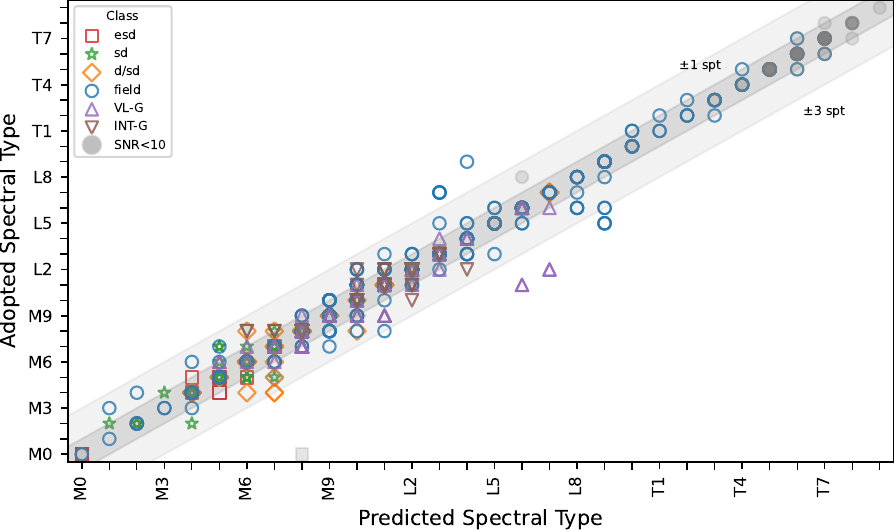}
    \caption{Same as Figure~\ref{fig:RF_scatter_actual_vs_predicted} for the KNN model.}
    \label{fig:KNN_scatter_actual_vs_predicted}
\end{figure*}

The multi-output KNN algorithm classifies spectral types at $\pm$1 SpT accuracy of $95.5\pm0.6\%$, and $\pm$3 SpT accuracy of $98.9\pm0.3\%$. The spectral type and class type classification result is presented in Figure \ref{fig:KNN_scatter_actual_vs_predicted} and \ref{fig:knn_dwarf_confusion_matrix}. A more detailed visualization of the model performance is shown in Appendix \ref{sec:Appendix_classification}, Figure \ref{fig:knn_SpT_confusion_matrix} for completeness. Compared to the other two models, KNN performs better at classifying d/sd and \textsc{vl-g} classes with a class accuracy of $89.5\pm0.9\%$, resulting in it being the best-performing model out of the three. We also plotted the distribution of SNR in our KNN prediction, shown in Figure \ref{fig:KNN_SNR}. Overall, \deleted{the higher SNR spectra tend to} \deleted{generate more accurate classifications.}\added{higher SNR did not seem to always lead to more accurate classification.} 

\subsection{Optimal SNR for Classification} \label{subsec:Optimal SNR for Classification}
\begin{figure}[htbp!]{}
    \centering
    \includegraphics[width=\columnwidth]
    {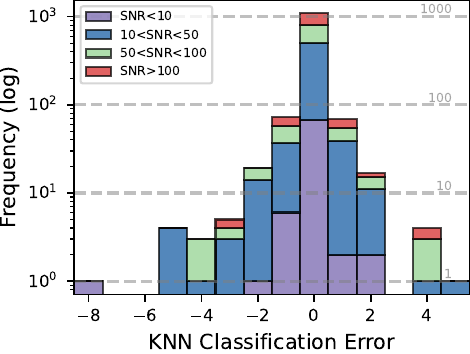}
    \caption{SNR histogram of the KNN model's predictions. The y-axis displays the log frequency, and the x-axis is the difference between the true label and the prediction (e.g., M0 = 0, L0 = 10; M0 $-$ L0 = $0-10$ would lead to a $-$10 classification error). Despite the expectation that high SNR spectra would be classified more accurately than low SNR spectra, \deleted{it is unclear why KNN classifications} \deleted{of high-SNR spectra do not} \deleted{replicate literature classifications.}\added{it is not obvious in our result. Instead, few high SNR objects were misclassified, likely because of the reasons explained in Section~\ref{subsec:Optimal SNR for Classification}. }}
    \label{fig:KNN_SNR}
\end{figure}

\begin{figure}[htbp!]{}
    \centering
    \includegraphics[width=\columnwidth]
    {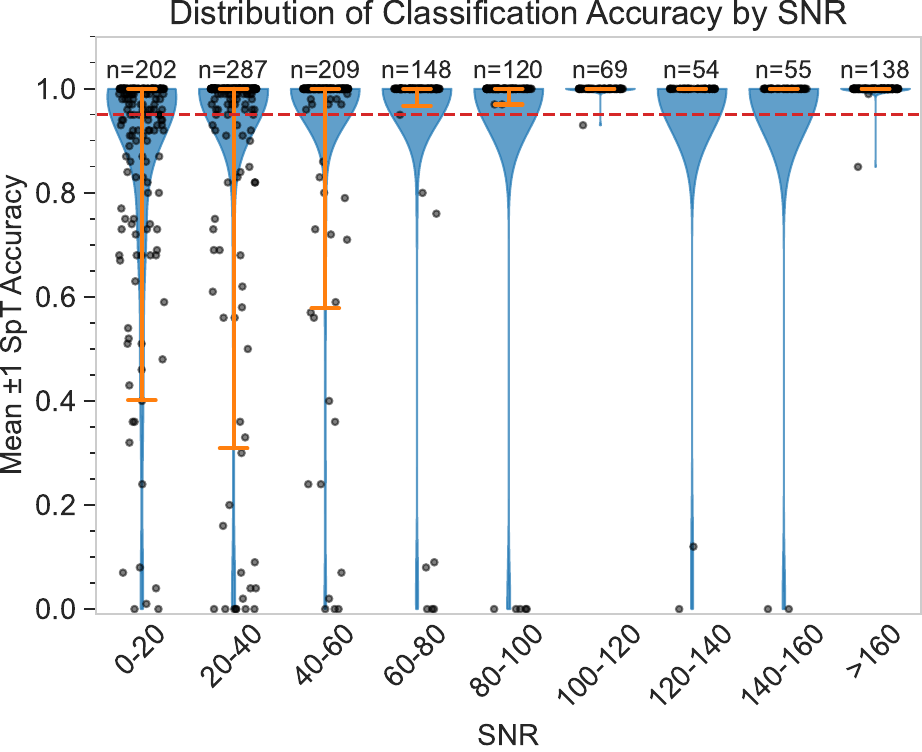}
    \caption{Mean $\pm$1 subtype classification accuracy versus SNR for our multi-output KNN classifier. Each data point (grey dot) is 100 synthetic spectra generated with Gaussian noise from table~\ref{tab:testset}. Each point shows the mean accuracy within a given SNR bin, with 5 to 95th percentiles indicated by the orange error bars. \added{The blue violin shapes describe the kernel-density distribution of accuracies for the 100 synthetic spectra in each SNR bin, illustrating the full spread of results.} Our result shows that classification accuracy improves with increasing SNR, and an SNR $\gtrsim$ 60 could guarantee robust classification.}
    \label{fig:SNR_Distribution}
\end{figure}

Next, we investigated the optimal SNR for spectral classification tasks. We generated 100 synthetic spectra per object in Table~\ref{tab:testset} by adding Gaussian‐distributed noise to the original spectrum, then classified the synthetic spectra using our KNN model, and finally binned the results by SNR to compute the mean classification accuracy, as is shown in Figure~\ref{fig:SNR_Distribution}, with numeric results provided in Table~\ref{tab:snr_accuracy}). We find a clear increase in the classification performance with SNR: the mean accuracy exceeds 95\% at SNR $\approx$ 50 and, at SNR $\approx$ 60, the lower (5th) percentile of the accuracy distribution already reaches 97\%. Thus, we recommend targeting SNR $\gtrsim$ 60 for new observations to guarantee robust classification of both spectral type and subtype. \added{The few outliers with SNR $\ge$ 120 that get misclassified are early M dwarfs, which peak at bluer wavelengths. This is likely because our model excludes wavelengths bluer than 0.89 microns, which reduces the ability to classify early M dwarfs. While Figure~\ref{fig:SNR_Distribution} demonstrates the model's robustness to random statistical noise, it suggests that the misclassification in Figure~\ref{fig:KNN_SNR} likely caused by physical properties not fully captured by our training standards.} Nevertheless, even in the lower quality spectra (SNR $<40$), the classifier attains a mean accuracy of $\approx90\%$, and over half of the spectra still achieve $\ge99\%$ accuracy, demonstrating that low-SNR data remain a useful resource for spectral typing. This result is consistent with the studies of spectral indices and ratios, which remain identifiable in low SNR data \citep{2005ApJ...623.1115C, 2010ApJS..190..100K, 2014ApJ...794..143B, 2024arXiv241101378B}.

\begin{figure}[htbp!]{}
    \centering
    \includegraphics[width=\columnwidth]
    {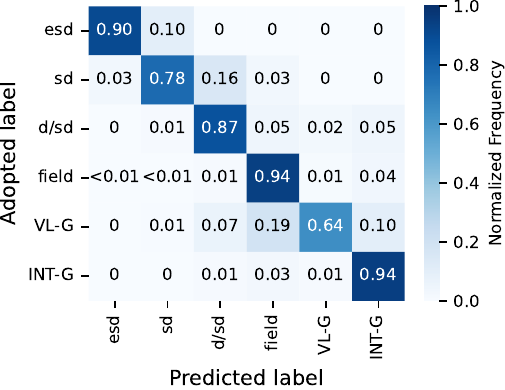}
    \caption{Same as Figure~\ref{fig:rf_dwarf_confusion_matrix} for the KNN model.}
    \label{fig:knn_dwarf_confusion_matrix}
\end{figure}

\begin{deluxetable}{l c c c}
\tablecaption{Classification Accuracy by SNR Bin\label{tab:snr_accuracy}}
\tablewidth{\textwidth}
\tablehead{
  \colhead{SNR Bin} 
& \colhead{$N$} 
& \colhead{Percentiles\textsuperscript{a}}
& \colhead{Mean}
}
\startdata
0–20    & 202 & [0.40, 0.99, 1.0] & 0.89 \\
20–40   & 287 & [0.31, 1.0, 1.0]  & 0.92 \\
40–60   & 209 & [0.58, 1.0, 1.0]  & 0.95 \\
60–80   & 148 & [0.97, 1.0, 1.0]  & 0.96 \\
80–100  & 120 & [0.97, 1.0, 1.0]  & 0.96 \\
100–120 &  69 & [1.0, 1.0, 1.0]   & 1.0  \\
120–140 &  54 & [1.0, 1.0, 1.0]   & 0.97 \\
140–160 &  55 & [1.0, 1.0, 1.0]   & 0.96 \\
$\ge$160& 138 & [1.0, 1.0, 1.0]   & 1.0  \\
\enddata
\tablenotetext{a}{5\%, 50\% (median) and 95\% classification‐accuracy percentiles.}
\tablecomments{Each array lists the 5\%, 50\% and 95\% percentiles across 100 synthetic‐spectra per spectrum in the test set. The Mean column gives the average classification accuracy across those spectra.}
\end{deluxetable}

\subsection{Algorithm Comparisons}
\label{subsection:Algorithm Comparisons}
Out of the three models tested, KNN performs the best in terms of spectral subtype and class accuracies. The performance of each model is shown in Table \ref{table:performance_metrics}.

\subsubsection{Spectral Subtype Classification}
\label{subsubsection:Spectral Subtype Classification}
\begin{figure*}[htbp]{}
    \centering
    \includegraphics[width=\textwidth]
    {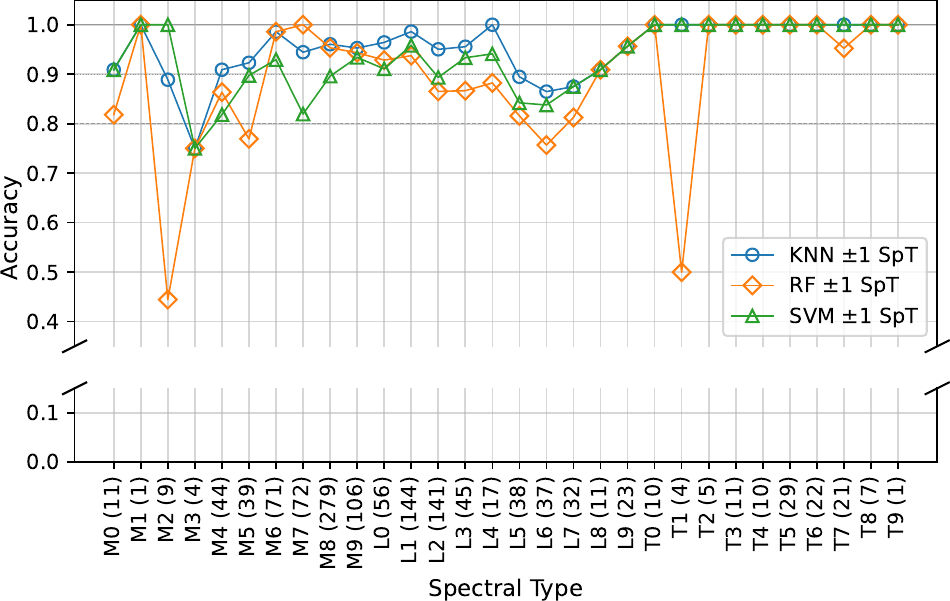}
    \caption{Model performance of KNN, RF, and SVM for each subtype. We show how accuracy varies for KNN (blue line and circles), RF (orange line and diamonds), and SVM (green line and triangles). The x-axis shows the subtypes from M0--T9, with the number in brackets representing the number of sources within that subtype in the testing set. The performance of all three models suggests L dwarfs might be harder to classify, compared to M and T dwarfs. }
    \label{fig:Model_comparison}
\end{figure*}

We compare all models' performances for each subtype and display it as M0--T9 subtype versus accuracy within $\pm$1 subtype in Figure \ref{fig:Model_comparison}. The number of test sources for each subtype label is displayed in brackets. In general, all models classify M and T dwarfs with significantly higher accuracy than L dwarfs. This discrepancy suggests that the reduced accuracy for L dwarfs arises primarily from the properties of the training set, as these sources exhibit greater intrinsic diversity likely due to cloud features in their photospheres. Conversely, the higher accuracy in T dwarf classification can be attributed to the absence of peculiar objects in the test set, as T dwarfs are predominantly field sources. Figures \ref{fig:rf_dwarf_confusion_matrix}, \ref{fig:svm_dwarf_confusion_matrix}, and \ref{fig:knn_dwarf_confusion_matrix} indicate that all our models perform optimally at classifying field dwarfs, with accuracies of 93\%, 92\%, and 94\% for RF, SVM, and KNN, respectively. This explains the high accuracy in T dwarf classifications, namely the test set lacks peculiar T dwarfs due to their intrinsic rareness, and our T dwarf sample is composed primarily of field sources.


\begin{deluxetable}{lccc}[htbp!]
\tablecaption{Performance Metrics for RF, SVM, and KNN Classifiers
\label{table:performance_metrics}}
\tablewidth{\textwidth}
\tablehead{ \colhead{Classifier}  & \colhead{Class}  & \colhead{SpT Accuracy} & \colhead{SpT Accuracy}\\
\colhead{}  & \colhead{Accuracy}  & \colhead{($\pm$1 subtype)} & \colhead{($\pm$3 subtype)}
}
\startdata
RF   & 86.15\%  & 91.54\% & 98.15\%  \\
SVM  & 82.69\%  & 91.08\% & 98.92\%  \\
KNN  & \textbf{89.46\%}  & \textbf{95.46\%} & \textbf{98.92\%}  
\enddata
\tablecomments{Best performing models are indicated with bold text.}
\end{deluxetable}

\subsubsection{Surface Gravity Classification}
\label{subsubsection:Surface Gravity Classification}

From the three class confusion matrices, it can be noted that our models classify field and metal-poor objects relatively well, but struggle at classifying very low surface gravity dwarfs. This also explains the trend in the classification scatter plot, where all three models perform relatively well on M and T dwarfs but struggle at classifying L dwarfs. This is likely due to the fact that most of our very low and intermediate surface gravity dwarfs are L dwarfs. Expanding the dataset with more low-gravity sources could provide a more comprehensive view of their spectral features, potentially leading to improved classification accuracy and a deeper understanding of these objects' physical properties. Interestingly, all our ML models appear to perform better at classifying \textsc{int-g} sources than \textsc{vl-g} sources\replaced{,}{.} \deleted{which are more different in terms} \deleted{of surface gravity.}\added{This is somewhat counter-intuitive, as \textsc{vl-g} objects represent a more significant deviation in surface gravity from the field population and might therefore be expected to be more easily distinguished.} The classification results of our three models are shown in Table~\ref{tab:Results}. Lastly, we \replaced{investigated}{investigate} gravity-sensitive wavelengths such as $K$-band and $H$-continuum \replaced{at}{in} \replaced{section}{Sections}~\ref{subsection:K-band Importance} and \ref{subsection:H-band Continuum}. 
\begin{deluxetable}{lcccc}
\tablecaption{Classification Results
\label{tab:Results}}
\tablewidth{\columnwidth}
\tablehead{
\colhead{Names} & \colhead{SpT$_\mathrm{RF}$} & \colhead{SpT$_\mathrm{SVM}$} & \colhead{SpT$_\mathrm{KNN}$} & \colhead{SNR}
}
\startdata
J00145014$-$0838231 & sdM9 & sdM9 & d/sdM7 & 31 \\
J01151621$+$3130061 & d/sdM8 & d/sdM8 & d/sdM9 & 28 \\
J03060166$-$0330590 & d/sdM9 & d/sdM9 & d/sdM9 & 52 \\
\multicolumn{5}{c}{\dots} \\
DENIS J23545990$-$1852210 & L1    & L1    & L1    & 118 \\
2MASSI J2356547$-$155310 & T5    & T5    & T5    & 18 \\
WISE J235716.49$+$122741.8 & T6    & T6    & T6    & 8 \\
\enddata
\tablecomments{\added{This table is available in its entirety in a machine-readable format. A portion is shown here for guidance regarding its form and content.} The references for this table are the same as in Table~\ref{tab:testset}.}
\end{deluxetable}

\subsection{Choice of Normalization and Scale Factor} 
\label{subsection:Choice of Normalization and Scale Factor}

Normalization in the $J$-band is commonly used for classifying UCDs, as this wavelength range provides a relatively stable pseudo-continuum with minimal broad molecular absorption compared to the $H$ and $K$ bands (e.g., \citealt{2010ApJS..190..100K,2022MNRAS.513..516F}). 
This allows for consistent flux calibration while preserving the diagnostic power of key spectral features in the $J$-band, such as the K \textsc{i} and Na \textsc{i} doublets. However, this normalization may suppress features outside the $J$-band that are critical for certain subclasses. For example, low-surface gravity dwarfs (\textsc{vl-g}) exhibit distinctive $H$-band continuum slopes and $K$-band spectral morphology \citep{2013ApJ...772...79A}, which are not emphasized in $J$-band-normalized spectra. This likely contributes to the poorer classification performance for \textsc{vl-g} objects seen in Figures \ref{fig:rf_dwarf_confusion_matrix}, \ref{fig:svm_dwarf_confusion_matrix}, and \ref{fig:knn_dwarf_confusion_matrix}, as the models are less sensitive to features which trace surface gravity outside the normalization window.

\section{Discussion} 
\label{sec:Discussion}

An overarching goal of this project was to create a framework for better classification of peculiar spectral types, namely the metallicity and gravity subclasses. Here we discuss our ability to classify each of these spectral subtypes in more detail\added{, and how spectral morphology affects the model's performance}.

\subsection{Surface Gravities}
\label{subsec:Surface Gravities}
This study's test set contains 138~$\gamma$ \deleted{(\textsc{vl-g})} and 209~$\beta$ \deleted{(\textsc{int-g})} sources.
We compared \replaced{10}{15} sources with literature spectral types that are both present in C18 and SPL. 
\replaced{All ten}{fourteen} sources were classified correctly within $\pm$1 subtype accuracy, achieving \replaced{100\%}{93\%} \replaced{overall}{SpT} accuracy. Moreover, all \replaced{ten}{fifteen} objects were \added{correctly} classified as exhibiting non-field surface gravity (either $\gamma$ or $\beta$). \replaced{Six out of ten}{Ten out of fifteen} objects exhibit consistent gravity-class classification, with \replaced{four}{five} \textsc{vl-g} ($\gamma$) objects misclassified as \textsc{int-g} ($\beta$) \deleted{despite having similar effective temperatures} \deleted{(base spectral type)} \added{but with predicted SpT within 1 subtype of the literature type}. This accuracy is largely consistent with our result shown in Figure~\ref{fig:knn_dwarf_confusion_matrix}, with \replaced{64\%}{62\%} accuracy in \textsc{vl-g} classification and \replaced{94\%}{100\%} accuracy in \textsc{int-g} classifications. The results are shown in Table~\ref{table:Surface Gravity Literature Comparison}. 

\begin{deluxetable*}{llccc}
\tablecaption{Surface Gravity Literature Comparison
\label{table:Surface Gravity Literature Comparison}}
\tablehead{
\colhead{Name} & \colhead{2MASS Designation} & \colhead{Predicted SpT} & \colhead{Literature SpT} & \colhead{References}
}
\startdata
2MASSW J0045214+163445 & J00452143+1634446 & L1$\gamma$ & L1.5$\gamma$ & 1, 2, 3 \\
2MASSI J0117474$-$340325 & J01174748$-$3403258 & L1$\gamma$ & L1$\gamma$ & 4, 5 \\
2MASS J02411151$-$0326587 & J02411151$-$0326587 & L0$\gamma$ & L0$\gamma$ & 6, 7, 8 \\
2MASS J03231004$-$4631263 & J03231004$-$4631237 & L1$\gamma$ & L0$\gamma$ & 6, 9, 1 \\
2MASS J10224821+5825453 & J10224821+5825453 & L2$\beta$ & L1$\beta$ & 6, 7, 1 \\
EROS-MP J0032$-$4405 & J00325584$-$4405058 & L0$\beta$ & L0$\gamma$ & 6, 7, 10 \\
2MASS J00550564+0134365 & J00550564+0134365 & L3$\beta$ & L2$\gamma$ & 9 \\
2MASS J02103857-3015313 & J02103857$-$3015313 & L0$\beta$ & L0$\gamma$ & 7, 11 \\
2MASSW J2206450$-$421721 & J22064498$-$4217208 & L3$\beta$ & L4$\gamma$ & 12, 2 \\
2MASS J15382417$-$1953116 & J15382417$-$1953116 & L4$\gamma$ & L4$\gamma$ & 7 \\
2MASS J17111353+2326333 & J17111353+2326333 & L0$\beta$ & L0$\gamma$ & 2, 8 \\
2MASS J16154255+4953211 & J16154255+4953211 & L6$\gamma$ & L4$\gamma$ & 7, 8 \\
2MASS J11544223$-$3400390 & J11544223$-$3400390 & L0$\beta$ & L0$\beta$ & 7, 13, 14 \\
2MASS J22134491$-$2136079 & J22134491$-$2136079 & L0$\gamma$ & L0$\gamma$ & 7, 8 \\
2MASS J15515237+0941148 & J15515237+0941148 & L2$\gamma$ & L3.5$\gamma$ & 7, 1 \\
\enddata
\tablecomments{The references are in the following order: spectral type reference, SpeX prism data reference, and discovery reference.}
\tablerefs{(1) \cite{2008AJ....136.1290R}; (2) \cite{2014ApJ...794..143B}; (3) \cite{2003IAUS..211..197W}; (4) \cite{2008ApJ...681..579B}; (5) \cite{2003AJ....126.2421C}; (6) \cite{2009AJ....137.3345C}; (7) \cite{2018AJ....155...34C}; (8) \cite{2007AJ....133..439C}; (9) \cite{2016ApJS..225...10F}; (10) \cite{1999AA...351L...5E}; (11) \cite{2015ApJS..219...33G}; (12) \cite{2000AJ....120..447K}; (13) \cite{2003AJ....126.1526B}; (14) \cite{2008ApJ...689.1295K}.}
\end{deluxetable*}

\subsection{Flux Variance Across Classes}
\label{subsec:Flux Variance Across Classes}
\begin{figure*}[ht!] 
\centering
\includegraphics[width=0.8\textwidth]{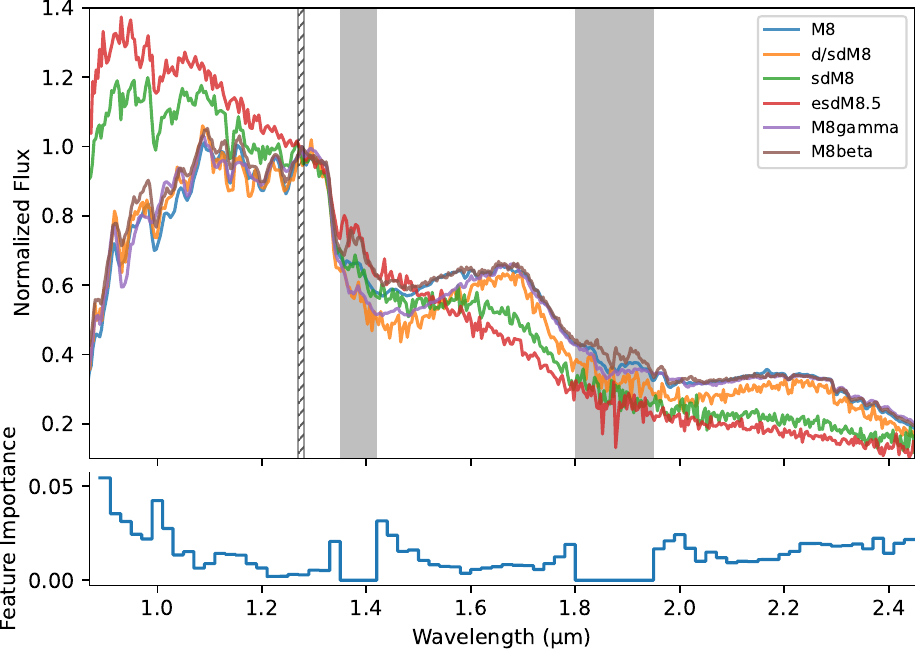} 

\caption{
    Comparison of spectral standards at fixed spectral types for M4 through L3, 
    showing field, low-gravity (\textsc{int-g}, \textsc{vl-g}), and low-metallicity 
    (d/sd, sd, esd) classes. Only spectral types with at least 3 different classes were selected. Normalization and feature importance are the same as Figure~\ref{fig:Combined_Spectra_featureImportance}. Note that all half-spectral types (e.g., esdM8.5) standards are treated as their integer spectral type (e.g., esdM8), as explained in Section~\ref{subsec:Spectral Standards}. These plots 
    illustrate that low-metallicity objects show the greatest variance in the 
    $zy$-band, corresponding to regions of high feature importance, while 
    low-gravity objects differ most in the $H$ and $K$ bands, where the overall feature importance is lower. 
    \textbf{The complete figure set (10 images) is available in the online journal.}
    \label{fig:spt_comparison}
}

\figsetstart
\figsetnum{19}
\figsettitle{Spectral Standards Comparison by Subtype}
\figsetgrpstart
\figsetgrpnum{19.1}
\figsetgrptitle{M4 Subtype Comparison}
\figsetplot{M4_fi_plot.pdf}
\figsetgrpnote{Comparison of field, low-gravity, and metal-poor spectral standards for the M4 subtype.}
\figsetgrpend
\figsetgrpstart
\figsetgrpnum{19.2}
\figsetgrptitle{M5 Subtype Comparison}
\figsetplot{M5_fi_plot.pdf}
\figsetgrpnote{Comparison of field, low-gravity, and metal-poor spectral standards for the M5 subtype.}

\figsetgrpend
\figsetgrpstart
\figsetgrpnum{19.3}
\figsetgrptitle{M6 Subtype Comparison}
\figsetplot{M6_fi_plot.pdf}
\figsetgrpnote{Comparison of field, low-gravity, and metal-poor spectral standards for the M6 subtype.}
\figsetgrpend

\figsetgrpstart
\figsetgrpnum{19.4}
\figsetgrptitle{M7 Subtype Comparison}
\figsetplot{M7_fi_plot.pdf}
\figsetgrpnote{Comparison of field, low-gravity, and metal-poor spectral standards for the M7 subtype.}
\figsetgrpend

\figsetgrpstart
\figsetgrpnum{19.5}
\figsetgrptitle{M8 Subtype Comparison}
\figsetplot{M8_fi_plot.pdf}
\figsetgrpnote{Comparison of field, low-gravity, and metal-poor spectral standards for the M8 subtype.}
\figsetgrpend

\figsetgrpstart
\figsetgrpnum{19.6}
\figsetgrptitle{M9 Subtype Comparison}
\figsetplot{M9_fi_plot.pdf}
\figsetgrpnote{Comparison of field, low-gravity, and metal-poor spectral standards for the M9 subtype.}
\figsetgrpend

\figsetgrpstart
\figsetgrpnum{19.7}
\figsetgrptitle{L0 Subtype Comparison}
\figsetplot{L0_fi_plot.pdf}
\figsetgrpnote{Comparison of field, low-gravity, and metal-poor spectral standards for the L0 subtype.}
\figsetgrpend

\figsetgrpstart
\figsetgrpnum{19.8}
\figsetgrptitle{L1 Subtype Comparison}
\figsetplot{L1_fi_plot.pdf}
\figsetgrpnote{Comparison of field, low-gravity, and metal-poor spectral standards for the L1 subtype.}
\figsetgrpend

\figsetgrpstart
\figsetgrpnum{19.9}
\figsetgrptitle{L2 Subtype Comparison}
\figsetplot{L2_fi_plot.pdf}
\figsetgrpnote{Comparison of field, low-gravity, and metal-poor spectral standards for the L2 subtype.}
\figsetgrpend

\figsetgrpstart
\figsetgrpnum{19.10}
\figsetgrptitle{L3 Subtype Comparison}
\figsetplot{L3_fi_plot.pdf}
\figsetgrpnote{Comparison of field, low-gravity, and metal-poor spectral standards for the L3 subtype.}
\figsetgrpend

\figsetend

\end{figure*}

\added{To better understand the classification of different gravity and metallicity classes, we examined the flux variance of objects with the same base spectral type. Figure~\ref{fig:spt_comparison} shows how metallicity and surface gravity affect its spectral morphology for each spectral type. The Figure set contains spectral types M4--L3, which were selected when each spectral type contains standards for at least 3 classes. At $zy$-band (0.9--1.1\,$\mu$m), the figure sets showed significant variances across metal-poor classes, and only small variances in low surface gravity objects. The J-band normalization method also appears to amplify the absolute variance in this bright portion of the spectrum, making these high-contrast features (e.g., metal-hydrides) dominant drivers of the classification. Since the model relies mostly on the $zy$-band, this explains the better performance of metallicity classification compared to gravity classification shown in Figure~\ref{fig:rf_dwarf_confusion_matrix}, \ref{fig:svm_dwarf_confusion_matrix}, and \ref{fig:knn_dwarf_confusion_matrix}. In contrast, the variance in $H$ and $K$ bands seems to be significant regardless of classes, and we therefore assess the impact of these bands by re-training the model without them.}

\subsection{K-band Importance}
\label{subsection:K-band Importance}

In order to assess the impact of the $K$-band on our gravity and metallicity classifications (see Section~\ref{subsection:Choice of Normalization and Scale Factor}), we retrained our machine‐learning models using a wavelength range restricted to 0.89--1.95\,$\mu$m, thereby excluding all $K$-band flux features. The $K$-band is known to harbor several important spectral features---including H$_2$ collision-induced absorption (CIA), \added{the 2.3~$\mu$m }CO\added{ bandhead}, and \added{the 2.2~$\mu$m }CH$_4$\added{ bandhead}---that are sensitive to atmospheric parameters. Based on current models of brown dwarf atmospheres, omitting the $K$-band is expected to reduce the accuracy of surface gravity determinations (due to the loss of H$_2$ CIA) and potentially affect metallicity classification (by removing the CO feature), with a possible impact on T dwarf classification via the CH$_4$ absorption feature.

\begin{figure*}[htbp]{}
    \centering
    \includegraphics[width=\textwidth]
    {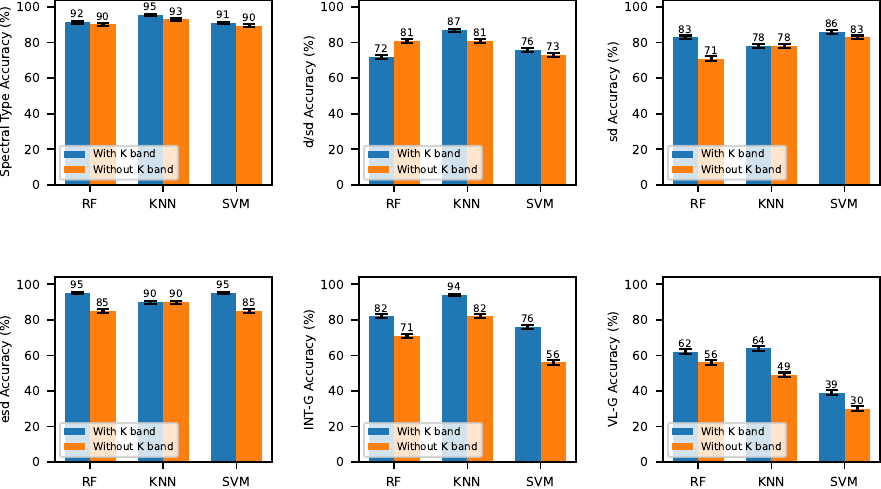}
    \caption{Grouped bar charts showing the performance before and after the removal of $K$-band. Accuracies of Spectral Type, d/sd, sd, esd, \textsc{int-g}, and \textsc{vl-g} are plotted against each ML model. Bars in blue correspond to models trained with $K$-band data, and those in orange correspond to models trained without. The figure illustrates that the exclusion of $K$-band significantly degrades gravity classification accuracy while exerting a smaller effect on metallicity diagnostics.}
    \label{fig:K_band_performance}
\end{figure*}

Figure~\ref{fig:K_band_performance} presents grouped bar charts comparing the classification accuracies (with binomial standard errors) for models trained with and without $K$-band data across six diagnostic tasks: spectral type, d/sd, sd, esd, \deleted{\textsc{vl-g}}\textsc{int-g}, and \textsc{vl-g}. Our analysis reveals that removal of the $K$-band leads to a significant degradation in gravity classification accuracy, with an average decrease of 14\% for \textsc{int-g} objects and 10\% for \textsc{vl-g} objects, which underscores the K-band’s importance in tracing surface gravity \citep{2013ApJ...772...79A}. This result also validates the importance of gravity-sensitive indices \added{in this region}, such as H2OD and H2O-2, suggested by \citet{2004ApJ...610.1045S} and \citep{2003ApJ...596..561M}.
In contrast, metallicity classification accuracy dropped by only about 3\% on average, suggesting that while CO absorption in the $K$-band contributes to metallicity diagnostics, other spectral features (e.g., FeH, CaH, TiO), especially those in $zy$-band, helped compensate for its absence.

Interestingly, the classification of T dwarfs remains unaffected, with all three models achieving at least 97\% accuracy even without $K$-band data. In the T dwarf classification scheme of \citet[][see Table 3]{burgasser:2006:1067}, five revised spectral indices are employed—H$_2$O-$J$, CH$_4$-$J$, H$_2$O-$H$, CH$_4$-$H$, and CH$_4$-$K$. Because four of these indices reside in the $J$ and $H$ bands, our results indicate that the CH$_4$ absorption at 2.2\,$\mu$m (i.e., the CH$_4$-$K$ index) does not play a critical role in T dwarf classification within our parameter space.
\subsection{H-band Continuum}
\label{subsection:H-band Continuum}
\begin{figure*}[htbp]{}
    \centering
    \includegraphics[width=\textwidth]
    {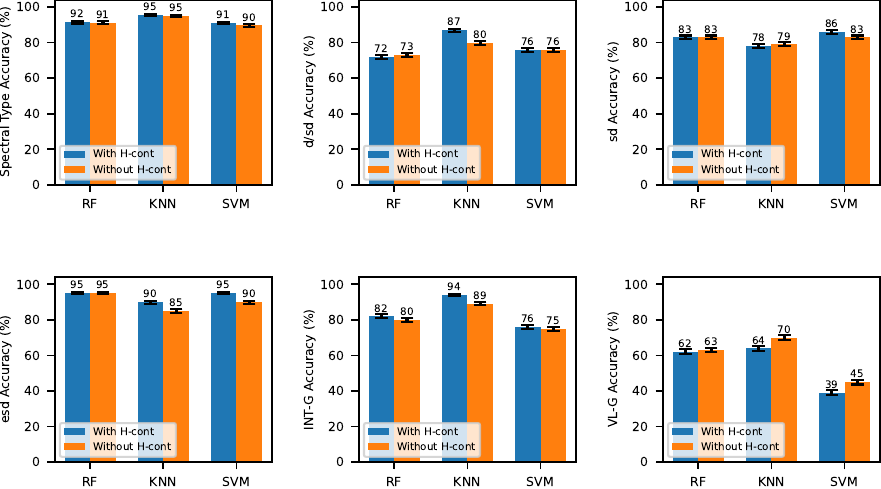}
    \caption{Same as Figure~\ref{fig:K_band_performance}, but for models trained with and without the $H$-band continuum features (1.47–1.67\,$\mu$m). In contrast to the $K$-band analysis, removal of the $H$-band continuum yields only minor changes in d/sd and gravity classification accuracy.}
    \label{fig:H_continuum_performance}
\end{figure*}

We further assess the influence of the gravity-sensitive, triangular-shaped $H$-band continuum on classification performance \citep{2013ApJ...772...79A}. In a manner analogous to the analysis described in Section~\ref{subsection:K-band Importance}, we removed all flux features in the wavelength range 1.47–1.67\,$\mu$m to isolate the effect of the $H$-band continuum morphology. Previous studies have noted that this distinctive continuum shape is not solely present in young, low-gravity objects but also manifests in d/sd “mild subdwarfs” \citep{2010ApJS..190..100K, 2016AJ....151...46A}. \added{The triangular H-band also manifests in red L dwarfs that are dusty but do not have any other indications of youth \citep{2013ApJ...772...79A}.} Figure~\ref{fig:H_continuum_performance} displays grouped bar charts analogous to Figure~\ref{fig:K_band_performance}, but for the $H$-band continuum analysis. The figure compares the accuracies (with corresponding standard errors) of models trained with and without $H$-band continuum features for the same six classification tasks. After the removal of the $H$-band continuum, our models exhibit an average decline of 2\% in d/sd classification and a slight increase of 1.7\% in gravity classification accuracy. \deleted{No significant correlation was observed between} \deleted{the $H$-band continuum and the} \deleted{classification of peculiarity.}\added{We find no significant correlation between the $H$-band continuum and peculiarity classification with our current methodology. We attribute this null result to our choice of normalization. Our primary method, normalizing to the $J$-band peak, was chosen because it is a standard technique that has proven robust for the general classification of M, L, and T dwarfs, which was the main goal of this work. However, as discussed in Section~\ref{subsection:Choice of Normalization and Scale Factor}, this normalization is not optimized to preserve broad continuum shapes like the triangular $H$-band feature. Therefore, this result does not rule out the $H$-band continuum as a powerful gravity diagnostic, but rather demonstrates that its effectiveness is highly dependent on the normalization scheme.}

\subsection{Subdwarfs}
\label{subsec:Subdwarfs}

In this study, our model classifies 163 M0--L1 metal-poor objects in the SPL. Although the model's parameter space includes peculiarities up to sdL3, we excluded sdL2 and sdL3 from further analysis due to insufficient testing data. The feature importance score (Figure~\ref{fig:Subdwarfs_with_Feature_Importances}) highlights the significant role of the FeH absorption (and possibly the lack of TiO absorption) features at 0.89 and 1.0\,$\mu$m, which are strong indicators of metallicity in M and L dwarfs. 

Interestingly, an increase in feature importance was observed immediately before and after the masked telluric bands at 1.4 and 1.9\,$\mu$m. This may suggest that the model is detecting parts of the water band, a critical indicator of metallicity \citep{2009ApJ...696..986C}. However, it is important to note that the utility of this feature is contingent on minimal airmass differences to mitigate contamination from Earth's atmosphere. 
We also compared sources with the literature spectral type listed in Table~\ref{table:subdwarf_spectra}. We excluded three duplicates and three unique sources that were randomly assigned to the validation set, leaving 15 sources for comparison. The model correctly classifies 12 out of 15 (80\%) metallicity prefixes, showing a good match between the model prediction and literature results. 
The model misclassified three mild subdwarfs (d/sd) as subdwarfs (sd), which suggests that the model correctly captures the peculiarity of the source, but fails to identify the strength of this peculiarity.  For the spectral subtype, the model correctly classifies 12 out of 15 sources within $\pm$1 subtype and 15 out of 15 within $\pm3$ subtypes.

\section{Summary, Conclusions, and Future Work} 
\label{sec:Conclusion}

In this study, we investigated the effectiveness of various machine-learning models in classifying spectral subtypes \added{and classes }(a proxy for $T_\mathrm{eff}$, $\log g$, and [M/H]) of M, L, and T dwarfs using NIR spectral data. We employed Random Forest\added{ (RF)}, K-Nearest Neighbors (KNN), and Support Vector Machine (SVM) algorithms, utilizing the mean flux within binned wavelength ranges as input features. Our training dataset comprised 70 spectral standards augmented with 1,000 synthetic spectra built from each standard\added{, for a total of 70,000 spectra}, while the test and validation sets included a diverse collection of field dwarfs, subdwarfs, and objects with varying surface gravities and metallicities from the Spex Prism Library.\added{ The test and validation sets consist of 1696 spectra.}

Our analysis focused on determining whether mean flux alone could effectively distinguish between different spectral types and peculiarities. We evaluated model performance across various spectral resolutions and assessed feature importance to identify critical wavelength regions influencing classification accuracy. We present the main findings of our study below.

\begin{enumerate}
    \item Model Performance: Among the tested algorithms, KNN outperformed others, achieving 95.5$\pm$0.6\% accuracy within $\pm$1 spectral subtype and 89.5$\pm$0.9\% accuracy in classifying class types \added{(surface gravity and metallicity)}. RF and SVM followed with 91.5$\pm$0.8\% \& 86.2$\pm$1.0\% and 91.1$\pm$0.8\% \& 82.7$\pm$1.0\% accuracies, respectively. All models maintained over 92\% accuracy in classifying field dwarfs.

    \item Feature Importance: 
    Spectral features (TiO, FeH) at 0.89 $\mu$m and 1.0 $\mu$m have the highest feature importance score, likely because they are sensitive to effective temperature, metallicity, and/or surface gravity \citep{2018AJ....155...34C}. Contrary to conventional spectral typing techniques \citep[e.g.,][]{2010ApJS..190..100K, 2018AJ....155...34C}, the $K$-band showed slightly higher importance than the $H$-band.

    \item We evaluated a range of spectral resolutions to assess their performance across several machine learning algorithms. The analysis indicates that a binning interval of $0.02\,\mu$m optimizes the classification of low-resolution NIR spectra.  However, despite its effectiveness, this binning approach may result in the loss of finer spectral details. Bins smaller than $0.02\,\mu$m could potentially retain information that larger bins might overlook. This could indicate that accurate spectral typing can be achieved with lower-resolution instruments. Future research could investigate the incorporation of additional features to further enhance classification accuracy.
    
    \item We investigated the influence of \added{the} $K$-band \added{(1.95--2.45$\mu$m)} and \added{the} $H$-\replaced{Continuum}{band continuum} (1.47--1.67$\mu$m) in our peculiarity classification. Our results indicate that the exclusion of \added{the} $K$-band decreases the gravity classification \added{accuracy} by around 12\% for \textsc{vl-g} and \textsc{int-g}, and an average decrease of 3\% for metallicity classification. However, for the exclusion of $H$-Continuum, we only found a decline of 2\% and 1.7\% for d/sd and surface gravity classification\added{, respectively}. These results support the critical role of $K$-band diagnostics in assessing surface gravity, while the $H$-band continuum exerts a comparatively limited influence on our classification model. \added{Metallicity classification was largely unaffected, suggesting that features in the $z$- and $y$-bands compensate for the absence of the $H$- and $K$-bands.}

    \item We explored the correlations between SNR and classification accuracies using synthetic spectra with Gaussian-distributed noise. We found that accuracy rises with SNR, surpassing the average accuracy of 95\% at SNR $\approx$ 50 and with the 5th percentile already above 97\% by SNR $\approx$ 60. As a result, \textit{we recommend \replaced{getting}{a target} SNR $\gtrsim$ 60 in future NIR observations for classification tasks}. We also note that even SNR $<40$ spectra still achieve an average accuracy of $\sim$ 90\%, in line with previous studies of \deleted{the} low SNR \replaced{index}{indices} \citep{2005ApJ...623.1115C, 2010ApJS..190..100K, 2013ApJ...772...79A, 2024arXiv241101378B}.
    
    \item We compared our predicted spectral types with those reported in the literature for various peculiarities. For \textsc{vl-g} and \textsc{int-g} classifications, we obtained literature spectral types from C18, as presented in Section~\ref{subsec:Surface Gravities}. Out of \replaced{10}{15} sources with literature spectral types, \deleted{all} \replaced{10}{14} sources were classified accurately within $\pm$1 subtype. All \replaced{10}{15} sources were correctly identified as low-surface gravity objects (either $\gamma$ or $\beta$), with replaced{5 out of 9}{8 out of 13} \textsc{vl-g} sources and \replaced{1 out of 1}{2 out of 2} \textsc{int-g} sources accurately classified in their respective gravity categories. This result is consistent with our test set classification performance, as depicted in Figure~\ref{fig:knn_dwarf_confusion_matrix}.

\end{enumerate}
Our model demonstrates strong capabilities in classifying both field dwarfs and mid-M to L dwarfs with peculiarity. However, the current classification model lacks continuity in certain peculiar subtypes, particularly among L-type subdwarfs and dwarfs with varying surface gravities shown in Table~\ref{tbl:standards}. Future work should incorporate additional spectral standards through new observations and employing statistical methods such as imputing missing values using mean or median flux to include previously masked flux features (e.g., wavelengths bluer than $0.89\,\mu$m), or extend similar analysis to bluer spectra such as those from Gaia. 

\added{Our training set is built upon synthetic spectra from single-epoch spectral standards, with each standard representing a "snapshot" of a star at one particular time. We acknowledge that this method does not explicitly model the time-domain spectral changes. The assumption in our model is that the chosen standards represent a typical or average state for their respective spectral types and that the Gaussian randomization around them can approximate small-scale physical variations in addition to observational noise. The primary limitation and potential bias of this approach is that the model's performance may be reduced when classifying an object observed in an extreme state of variability. A solution and important future work to this is to obtain multi-epoch observations for known variable stars, which can create even more powerful and physically grounded classification models.}

\added{The models presented in this work were trained and optimized for low-resolution (R$\sim$120) NIR spectra from the SpeX instrument. However, the underlying framework---using an ML classifier on binned spectral flux---is broadly applicable and can be adapted for data from other low-resolution NIR spectrographs (e.g., JWST NIRSpec, SOAR TripleSpec4.1). We caution that the SpeX-trained model itself is not directly transferable, as it is sensitive to the specific instrumental profile and noise characteristics of its training data. The most scientifically rigorous application of this framework to a new instrument would therefore involve retraining the classifier on a representative set of standards from that facility. A valuable future work to test the generalizability of the method is to evaluate downsampled higher-resolution spectra from different surveys or retrain the model based on observations from the target instrument.}

\begin{acknowledgments} 
\label{sec:acknowledgments}

This research was supported by the Triton Research \& Experiential Learning Scholars (TRELS) program and the Summer Training Academy for Research Success (STARS) program at the University of California, San Diego, which provided stipends for this study. Funding for this work was also provided by NASA (grant number 80NSSC24K0156).

Visiting Astronomer at the Infrared Telescope Facility, which is operated by the University of Hawaii under contract 80HQTR24DA010 with the National Aeronautics and Space Administration.

This publication makes use of data products from the Two Micron All Sky Survey, which is a joint project of the University of Massachusetts and the Infrared Processing and Analysis Center/California Institute of Technology, funded by the National Aeronautics and Space Administration and the National Science Foundation

The authors wish to recognize and acknowledge the very significant cultural role and reverence that the summit of Maunakea has always had within the indigenous Hawaiian community. We are most fortunate to have the opportunity to conduct observations from this mountain.
\end{acknowledgments}

\section*{Data Availability}
\label{sec:Data_Availability}
The data used for testing are available in the SpeX Prism Library (SPL) at {\footnotesize\url{https://github.com/aburgasser/splat}}. All models and pre-processed data used for training and testing are available at {\footnotesize\url{https://github.com/StellarDataLab/SpecAI}}.

\vspace{5mm}
\facilities{IRTF (SpeX)}


\software{\textsc{astropy} \citep{2013A&A...558A..33A,2018AJ....156..123A, 2022ApJ...935..167A},  \textsc{Matplotlib} \citep{ Hunter:2007}, \textsc{SPLAT} \citep{2017ASInC..14....7B}, \textsc{Numpy} \citep{harris2020array}, \textsc{Scipy} \citep{2020SciPy-NMeth}, \textsc{Scikit-Learn} \citep{scikit-learn}}



\appendix
\section{Classification Performance Across Models}
\label{sec:Appendix_classification}
\begin{figure*}[ht]{}
    \centering
    \includegraphics[width=\textwidth]
    {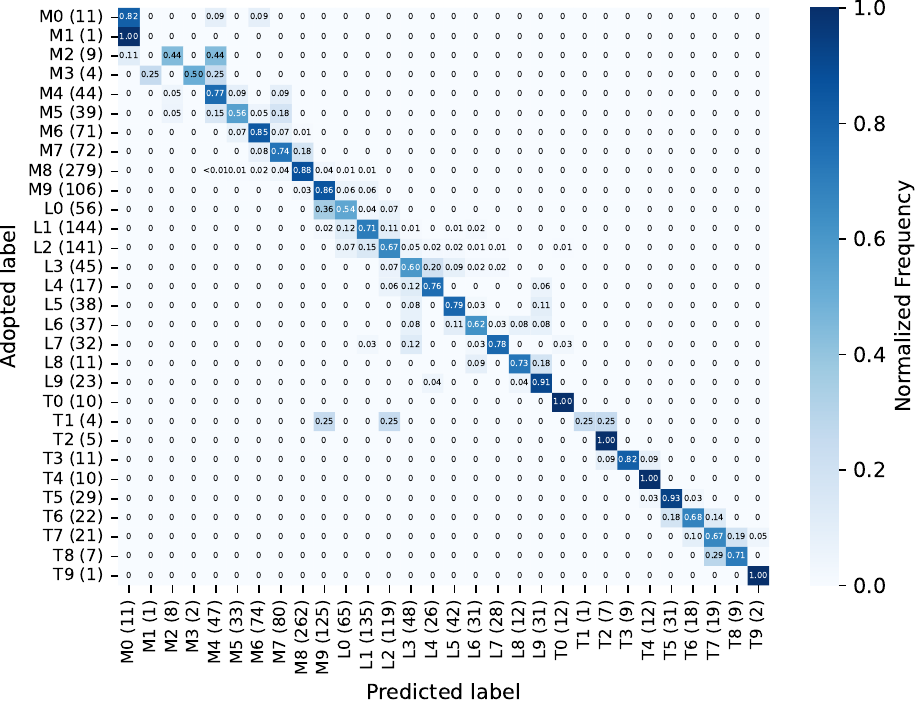}
    \caption{Confusion Matrix for the Random Forest (RF) model in classifying spectral subtypes. All field and peculiar subtypes are included and grouped under their base type (i.e., L6$\gamma$ goes under L6). The values are normalized across each row (actual label), i.e., each row sums to 1. This graph demonstrates the predicted label of the RF model versus the actual label defined in the dataset.}
    \label{fig:rf_SpT_confusion_matrix}
\end{figure*}

\begin{figure*}[ht]{}
    \centering
    \includegraphics[width=\textwidth]
    {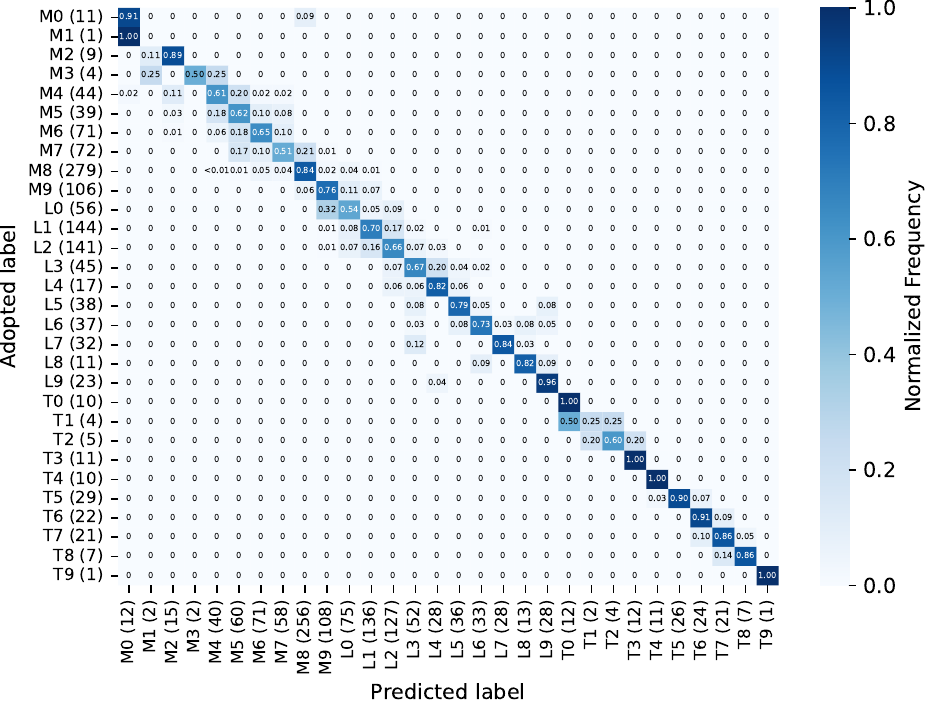}
    \caption{Same as Figure~\ref{fig:rf_SpT_confusion_matrix}, but for SVM.}
    \label{fig:svm_SpT_confusion_matrix}
\end{figure*}

\begin{figure*}[ht]{}
    \centering
    \includegraphics[width=\textwidth]
    {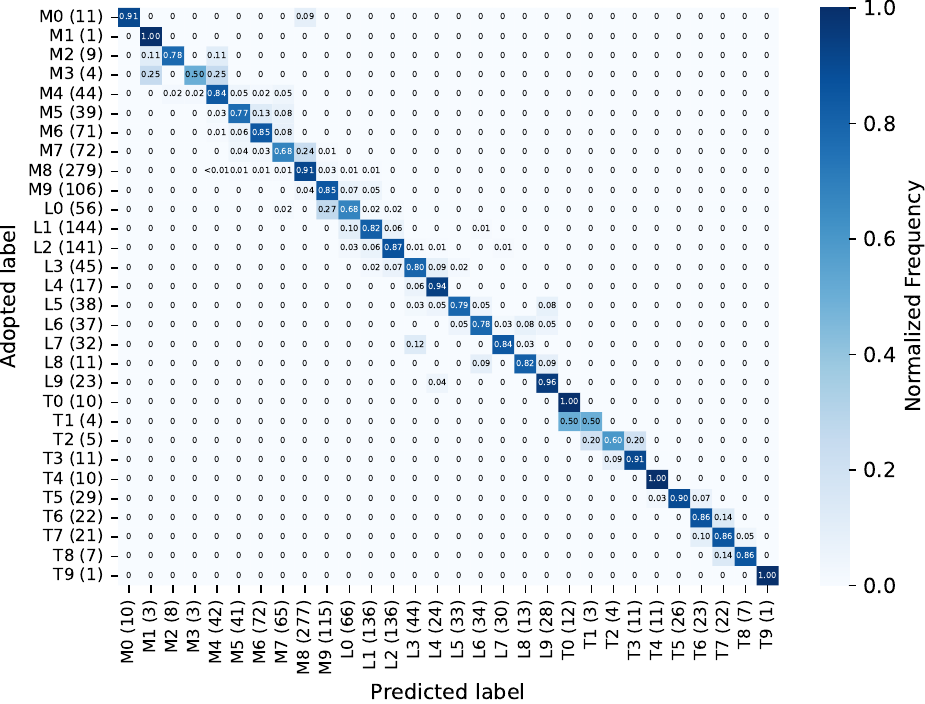}
    \caption{Same as Figure~\ref{fig:rf_SpT_confusion_matrix}, but for KNN.}
    \label{fig:knn_SpT_confusion_matrix}
\end{figure*}

We present the performance of three machine learning models—Random Forest (RF), k-Nearest Neighbors (KNN), and Support Vector Machine (SVM)—in classifying spectral types M0--T9. For each model, confusion matrices are used to compare the adopted labels with the predicted labels. In these matrices, the diagonal elements represent the fraction of correct predictions, and the off-diagonal elements indicate misclassifications. Each row is normalized so that its elements sum to 1. We evaluated the classification accuracy across spectral subtypes for each model using confusion matrices. 
Figure~\ref{fig:rf_SpT_confusion_matrix} shows that the RF model performs well for late M and T dwarfs but struggles with early M and L dwarfs. In contrast, the SVM model (Figure~\ref{fig:svm_SpT_confusion_matrix}) exhibits a similar trend with notably reduced scattering for early M dwarfs and a slight improvement for L dwarfs. The KNN model (Figure~\ref{fig:knn_SpT_confusion_matrix}) further minimizes scattering for early M and L dwarfs relative to the RF and SVM model. Detailed analysis of these models can be found in Section~\ref{sec:results} and \ref{sec:Discussion}.

\section{Optimal Flux Binning}
\label{sec:Appendix_binning}
Flux binning is the primary approach for determining the feature size and is critical to our model's performance. Following the methodology described in Section~\ref{subsubsec:Building the Training set}, we constructed training sets using wavelength bins ranging from $0.01,\mu$m to $0.04,\mu$m and included an additional bin width of $0.10,\mu$m for comparison.  We determine the optimal wavelength binning intervals by applying a simple random forest classifier (with \texttt{n\_estimator} = 100, \texttt{max$\_$depth} = None, \texttt{min$\_$samples$\_$split} = 2, \texttt{min$\_$samples$\_$leaf} = 1, and \texttt{bootstrap} = True) to the validation set described in Section~\ref{sec:Testing and Validation set}. 

Table~\ref{table:performance_metrics_of_different_bins} summarizes the resulting classification accuracies and the corresponding bin sizes. Our analysis indicates that a bin width of $0.02\,\mu$m optimally balances model performance with the preservation of spectral detail. For context, the native resolution of the spectra---defined as the median spacing between adjacent wavelength points---is approximately $0.003\,\mu$m. Although the $0.02\,\mu$m bin size appears to represent a local minimum, our parameter space was not exhaustively explored; future studies may further refine the optimal binning strategy for spectral downsampling.

\begin{deluxetable}{ccc}
\tablecaption{Performance Metrics For Varying Wavelength Bins\label{table:performance_metrics_of_different_bins}}
\tablewidth{\columnwidth}
\tablehead{
\colhead{Bins ($\mu$m)}  & \colhead{Class Accuracy (\%)} & \colhead{SpT Accuracy (\%)} }
\startdata
0.01 & 91.67\%  & 82.58\%  \\
0.02 & \textbf{90.66\%}  & \textbf{84.85\%} \\
0.03 & 90.66\%  & 84.09\% \\
0.04 & 90.66\%  & 82.58\% \\
0.10 & 89.39\%  & 82.83\%  \\
\enddata
\end{deluxetable}

\bibliography{references.bib}{}
\bibliographystyle{aasjournal}



\end{document}